\documentclass[aps, prb, twocolumn, amsmath, amssymb, longbibliography]{revtex4-2}

\usepackage{booktabs}
\usepackage[caption=false]{subfig}
\usepackage{graphicx}
\graphicspath{{figures/}}
\usepackage{tikz}
\usetikzlibrary{shapes}
\usepackage{braket}
\usepackage[colorlinks=true,linkcolor=blue,citecolor=blue,breaklinks]{hyperref}
\usepackage{mathtools}
\newcommand{\figref}[1]{Fig.~\ref{#1}}
\newcommand{\tabref}[1]{Table \ref{#1}}
\newcommand{\secref}[1]{Section \ref{#1}}
\newcommand{\refref}[1]{Ref.~\cite{#1}}
\newcommand{\equref}[1]{Eq.~\eqref{#1}}
\newcommand{\appref}[1]{Appendix~\ref{#1}}

\newcommand{\vect}[1]{{\textbf{#1}}}
\newcommand{\R}{\vect{r}}
\newcommand{\h}{h}

\DeclareMathOperator{\sign}{sgn}

\tikzset{octagon/.style ={shape=regular polygon, regular polygon sides=8, draw, sharp corners}}
\tikzset{dodecagon/.style ={shape=regular polygon, regular polygon sides=12, draw, sharp corners}}
\tikzset{hexagon/.style = {shape=regular polygon, regular polygon sides = 6, draw, sharp corners}}
\tikzset{triangle/.style ={shape=regular polygon, regular polygon sides=3, draw, sharp corners}}

\newcommand{\hexagon}[1]{\begin{tikzpicture}\node[hexagon, scale = #1]{};\end{tikzpicture}}
\newcommand{\octagon}[1]{\begin{tikzpicture}\node[octagon, scale  = #1]{};\end{tikzpicture}}
\newcommand{\dodecagon}[1]{\begin{tikzpicture}\node[dodecagon, scale = #1]{};\end{tikzpicture}}

\begin{document}

	\title{Field stability of Majorana spin liquids in antiferromagnetic Kitaev models}
	\author{Christoph Berke}
	\email{berke@thp.uni-koeln.de}
	\author{Simon Trebst}	
	\author{Ciar\'an Hickey}
	
	\affiliation{Institute for Theoretical Physics, University of Cologne, 50937 Cologne, Germany}
	\date{\today}
	
	\begin{abstract}
	Magnetic fields can give rise to a plethora of phenomena in Kitaev spin systems, such as the formation of non-trivial spin liquids in two and three spatial dimensions. 
	For the original honeycomb Kitaev model, 
	it has recently been observed that the sign of the bond-directional exchange is of crucial relevance for 
	the field-induced physics, with antiferromagnetic couplings giving rise to an intermediate spin liquid
	regime between the low-field gapped Kitaev spin liquid and the high-field polarized state, which is not
	present in the ferromagnetically coupled model.
	Here, by employing a Majorana mean-field approach for a magnetic field pointing along the [001] direction, we present a systematic study of field-induced spin liquid phases for a variety of two and 
	three-dimensional lattice geometries.
	We find that antiferromagnetic couplings generically lead to 
	(i) spin liquid phases that are considerably more stable in field than those for ferromagnetic couplings,
	and (ii) 
	an intermediate spin liquid phase which arises from a change in the topology 
	of the Majorana band structure.
	Close inspection of the mean-field parameters reveal that the intermediate phase occurs due to a field-driven 
	sign change in an `effective' $z$-bond energy parameter. Our results clearly demonstrate the richness of the Majorana physics of the antiferromagnetic Kitaev models, in comparison to their ferromagnetic counterparts. 
	\end{abstract}
	
	\maketitle


	\section{Introduction}
	Few fields of condensed matter physics have experienced a similar flourishing in recent decades as the study of quantum spin liquids (QSLs). These are captivating states of matter, in which strongly interacting spins avoid symmetry breaking magnetic order down to the lowest temperatures due to strong quantum fluctuations \cite{savary_quantum_2016,balents_spin_2010, knolle_field_2019,anderson_resonating_1973}. A powerful framework to understand such states has been to express -- or \textit{fractionalize} -- the constituent spins in terms of emergent degrees of freedom, similar to spin-charge separation in Luttinger liquids \cite{haldane_textquotesingleluttinger_1981}. In the context of QSLs, the spins naturally decompose into spinons, either bosons or fermions, and emergent gauge fields in a deconfined phase \cite{broholm_quantum_2020, zhou_quantum_2017,wen_quantum_2007,read_large-n_1991,senthil_$z_2$_2000}.  
	
	A paradigmatic example of a QSL occurs in Kitaev's eponymous honeycomb model \cite{kitaev_anyons_2006}, 	which provides an analytically tractable example of a local quadratic spin Hamiltonian whose \textit{exact} ground state is a QSL. In this model, $S=1/2$ spins are subject to strong exchange frustration arising from bond-directional Ising interactions. Its exact
	solution is typically formulated in terms of itinerant Majorana fermions and a static $\mathbb{Z}_2$ gauge field, which has allowed for 
	a plethora of conceptual studies of spin liquid physics phenomena \cite{hermanns_physics_2018}. The significance of this model, however, goes far beyond being the drosophila of QSLs: Despite its 
	seemingly artificial interactions,  Khaliullin, Jackeli, and Chaloupka have shown in a series of works 
	\cite{khaliullin_orbital_2005,jackeli_mott_2009,chaloupka_kitaev-heisenberg_2010} that bond-directional Kitaev-like interactions can actually be realized in spin-orbit entangled $j\!=\!1/2$ Mott insulators \cite{witczak-krempa_correlated_2014, kim_novel_2008, kim_phase-sensitive_2009}. 
	Over the last decade, a variety of such $4d$ and $5d$ Kitaev materials have been synthesized and subject of an intense experimental
	search for fingerprints of spin liquid physics \cite{trebst_kitaev_2017, takagi_concept_2019}. Prominent examples include the honeycomb materials Na$_2$IrO$_3$ \cite{singh_antiferromagnetic_2010,chunDirectEvidenceDominant2015},
	Li$_2$IrO$_3$ \cite{singh_relevance_2012}, and $\alpha$-RuCl \cite{plumb_rucl_2014, banerjee_proximate_2016, banerjeeNeutronScatteringProximate2017, majumder_anisotropic_2015}, all of which have been identified as having ferromagnetic (FM) Kitaev interactions \cite{winter_challenges_2016, winter_models_2017}.
	However, all of these materials exhibit magnetically ordered states at low temperatures \cite{chaloupka_kitaev-heisenberg_2010,kimchi_kitaev-heisenberg-$j_2$-$j_3$_2011,chaloupka_zigzag_2013,rau_generic_2014,rousochatzakis_phase_2015,janssen_magnetization_2017, kitagawa_spinorbital-entangled_2018, sears_magnetic_2015,  kubota_successive_2015}, indicating the presence of additional interactions, beyond those of the pure Kitaev model, which tend to stabilize conventional magnetic ordering.

	Motivated by the discovery that a magnetic field suppresses the magnetic order in $\alpha$-RuCl and brings it to the proximity of a potential QSL phase \cite{johnson_monoclinic_2015, kubota_successive_2015, majumder_anisotropic_2015, sears_phase_2017, hirobe_magnetic_2017, kasahara_unusual_2018, banerjee_excitations_2018, jansa_observation_2018, wolter_field-induced_2017,kasahara_majorana_2018}, intensive theoretical research on the Kitaev model in an external field has been initiated \cite{hickey_emergence_2019, jiang_possible_2011, gohlke_dynamical_2018, fey_sebastian_field_nodate,jiang_field_2018,zou_field-induced_2019,liang_intermediate_2018,jiang_possible_2011,ronquillo_signatures_2019,nasu_successive_2018,janssen_heisenbergkitaev_2019}.
	As already shown by Kitaev himself, a weak magnetic field in the perturbative regime induces next nearest neighbor hopping between the Majorana fermions and gaps out the otherwise gapless spectrum.
	Going beyond the perturbative regime, recent studies found evidence that the antiferromagnetic (AFM) Kitaev model hosts a gapless QSL with a $U(1)$ gauge structure at intermediate field strengths, using numerical exact diagonalization \cite{hickey_emergence_2019} and density matrix renormalization group techniques \cite{jiang_field_2018}.
	Shortly afterwards, implementations of such AFM interactions in actual materials were proposed, e.g. in $4f$ electron based systems \cite{jang_antiferromagnetic_2019, motome_materials_2020, sugita_antiferromagnetic_2019}, further enriching the zoo of possible Kitaev materials. 
	
	In this manuscript, we discuss the physics of the Kitaev model in a [001] magnetic field 
	on various two- and three-dimensional lattice geometries and compare and contrast the FM and AFM cases. 
	Technically, we adopt a Majorana mean-field approach that was first applied by \citet{nasu_successive_2018} and which focuses on the Majorana signatures of the model. Fractionalization into these quasiparticles is the defining property of the Kitaev model and believed to be the driving force for a number of astonishing experimental findings \cite{sandilands_scattering_2015, sandilands_spin-orbit_2016, nasu_fermionic_2016, banerjeeNeutronScatteringProximate2017, banerjee_proximate_2016,kasahara_majorana_2018, kasahara_unusual_2018}, making a Majorana-focused point of view a well-motivated starting point.  
	We show that, in all cases, the Kitaev spin liquid (KSL) in the AFM Kitaev model is significantly more stable when ramping up the magnetic field than the one for the analogous FM model. Furthermore, for AFM couplings, the models generically host an intermediate spin liquid phase, between the KSL and the trivial field-polarized phase, whose precise nature depends on the underlying lattice structure.


\begin{figure}
	\centering 
	\hspace*{-0.4cm}
	\begin{tikzpicture}
	\node[inner sep = 0pt] {\includegraphics{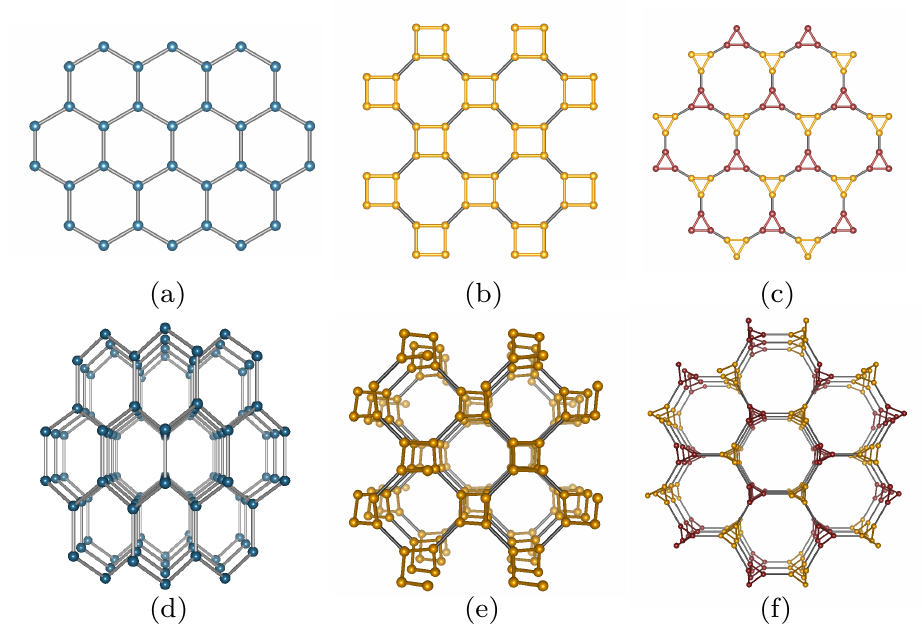}};
	\node[opacity=0]{\subfloat[][\label{vesta:HC}]{}};\node[opacity=0]{\subfloat[][\label{vesta:SO}]{}};\node[opacity=0]{\subfloat[][\label{vesta:YK}]{}};
	\node[opacity=0]{\subfloat[][\label{vesta:103b}]{}};\node[opacity=0]{\subfloat[][\label{vesta:103a}]{}};\node[opacity=0]{\subfloat[][\label{vesta:83b}]{}};	
	\end{tikzpicture}
	\caption{\textbf{Overview over tricoordinated lattices} in 2D, \protect\subref{vesta:HC} honeycomb lattice, \protect\subref{vesta:SO} square-octagon lattice, \protect\subref{vesta:YK} decorated honeycomb lattice), and their 3D analogues, \protect\subref{vesta:103b} (10,3)b or hyperhoneycomb lattice, \protect\subref{vesta:103a} (10,3)a or hyperoctagon lattice, \protect\subref{vesta:83b} (8,3)b lattice.}
	\label{fig:lats}
\end{figure}
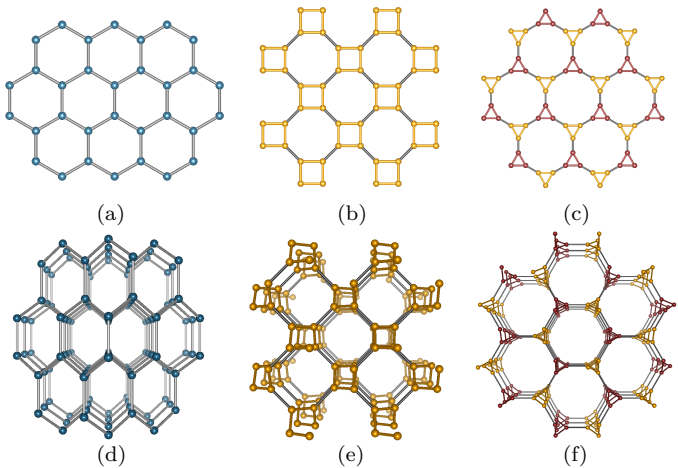

\subsection*{\label{sec:Overview} Overview of results}
As essential ingredients for this study, we have chosen three two-dimensional (2D) and three three-dimensional (3D) tricoordinated lattice geometries, as illustrated in Fig.~\ref{fig:lats}.
The distinguishing feature of the QSL ground state of the Kitaev model for each of these lattice geometries is the resulting band structure of the 
Majorana fermions, whose features (gapped versus gapless, topological versus trivial) depend sensitively on the underlying lattices.
An overview of the lattices, together with information on their zero-field band structure, is given in \tabref{table:summary}. For the 2D models, we discuss a gapped and topologically trivial Kitaev model on the square-octagon lattice \cite{yang_mosaic_2007} and a gapped and topologically non-trivial Kitaev model on the decorated honeycomb lattice, whose ground state is a chiral spin liquid \cite{yao_exact_2007}. We also briefly recapitulate the gapless Kitaev model on the honeycomb lattice, already discussed within this framework in \refref{nasu_fermionic_2016}. For the 3D models, we discuss one example for each codimension $d_c=1,2,3$ of the Majorana Fermi surface \footnote{In the following, the terminus `Fermi surface' refers to the manifold of gapless excitations, independent of dimension and codimension.}: The lattice (8,3)b \footnote{We use the so-called Schl{\"a}fli symbol $(p,c)$ to label the lattices. $p$ is the elementary loop length and $c\!=\!3$ refers to the tricoordination.} that hosts Weyl points ($d_c\!=\!3$), the hyperhoneycomb lattice that hosts a Majorana Fermi line ($d_c\!=\!2$) and the hyperoctagon lattice which exhibits a full Fermi surface ($d_c\!=\!1$). As shown in \figref{fig:lats} these three particular lattices are the natural 3D analogues of the three 2D lattices. 

 
\begin{figure}[b]
	\begin{tikzpicture}
		\node[inner sep=0pt]{\includegraphics[scale=1]{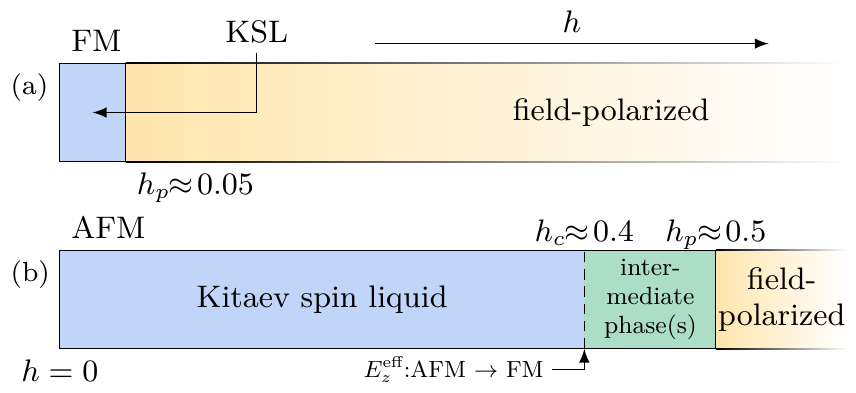}};
		\node[opacity=0]{\subfloat[][\label{genericPD_FM}]{}};\node[opacity=0]{\subfloat[][\label{genericPD_AFM}]{}};
	\end{tikzpicture}
	\caption{\textbf{Generic phase diagram} for  \protect\subref{genericPD_FM} the  FM and  \protect\subref{genericPD_AFM} the AFM Kitaev model under a magnetic field in [001] direction. The dashed line denotes a change in the topology of the Fermi surface and solid lines the transition to the field-polarized phase. 
	$E_z^{\text{eff}}(h)$ changes sign at $h_c$, while the nearest neighbor correlation function $\langle S_i^z S_j^z\rangle_z$ changes sign within the intermediate phase.}
	\label{fig:genericPD}
\end{figure}

\setlength{\tabcolsep}{9pt}
\begin{table*}\centering
	\caption{\textbf{Overview of two and three dimensional tricoordinated lattices} and the results for the critical [001] fields at which the polarized phase is entered ($h_p$), and for the fields where the non-universal intermediate phase sets in ($h_{c}$) for AFM couplings. For the decorated honeycomb and (8,3)b lattice, $h_{c}$ gives the first critical field strength.}
	\label{table:summary}
	\begin{tabular}{llllll}
		\toprule 
		&&FM & \multicolumn{2}{c}{AFM} \\
		\cmidrule{4-5}
		lattice & band structure at $\h=0$ & $h_p$ &$h_{c}$ & $h_{p}$  & Fermi surface at $h_c$  \\
		\midrule 		
		square-octagon & gapped ($\nu= 0$) & 0.075 & 0.423 & 0.527 & nodal line \\
		decorated honeycomb & gapped ($\nu=\pm1$) & 0.025 & 0.405 & 0.523 & Dirac points/nodal line \\
		honeycomb & Dirac cones & 0.042 & 0.417 & 0.503 & nodal line\\
		\cmidrule{1-6}
		(8,3)b & Weyl nodes ($d_c=3$) & 0.030 & 0.415 & 0.489 & nodal line \\
		(10,3)b / hyperhoneycomb & nodal line ($d_c=2$) &  0.028 & 0.415 & 0.485 & nodal line \\ 
		(10,3)a / hyperoctagon & Fermi surface ($d_c=1$) & 0.028 & 0.416 & 0.487 & flat bands \\		
		\bottomrule 
	\end{tabular}
\end{table*}

Switching on a field $\h$ in the [001] direction and using a Majorana mean-field approach, we find the generic behavior summarized in \figref{fig:genericPD} for all lattices under consideration. The phase diagram of the FM Kitaev model (\figref{genericPD_FM}) shows a single transition from the KSL phase to the field-polarized phase at a critical field strength $h_p\approx 0.05$.
As \figref{genericPD_AFM} depicts, the KSL with AFM couplings is stable to much higher fields and the polarized phase only appears at $h_p\approx 0.5$. A universal feature for all lattices is the emergence of an additional intermediate phase(s) in the AFM model, appearing at field strengths of $h_c \approx 0.4$. The precise critical fields for the FM and AFM models are summarized in \tabref{table:summary}. 

\begin{figure}[b]
	\centering 
	\includegraphics[scale=1]{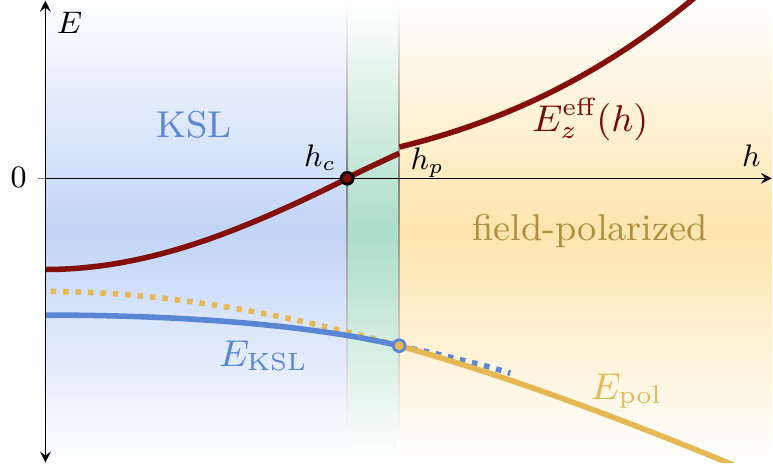}
	\caption{\textbf{Schematic representation} of the evolution of $E_z^\text{eff}$ and the energies in the KSL ($E_{\text{KSL}}$) and polarized phase ($E_{\text{pol}}$).}
	\label{fig:energies}
\end{figure}

In the presence of a [001] field the Kitaev model can be written as
\begin{align}
\notag H &=  \sum_\gamma H_\gamma + H_h \\
 &= -  \sum \limits_{\langle ij \rangle_\gamma} J_\gamma S^\gamma_i S^\gamma_j - \h \sum \limits_i S^z_i,
\label{eq:KitaevHam}
\end{align}
where $J_\gamma>0$ ($J_\gamma<0$) is the FM (AFM) Kitaev coupling along the bond $\gamma \in \lbrace x,y,z \rbrace$. For the AFM model, with isotropic couplings $J_\gamma = J$, the nature of both the transition at $h_c$ and the intermediate phase(s) between $h_c$ and $h_p$ can in fact be directly related to the nature of the corresponding zero-field Kitaev model with variable $J_z$ and fixed AFM $J_x$, $J_y$. 
Indeed, by defining an effective parameter 
\begin{equation}
	E_z^\text{eff}(h) = \braket{H_z^{\text{eff}}(h)}/N = \braket{ H_z + H_h }/N - h^2/2J \, ,
	\label{eq:Ezeff}
\end{equation}
we find for all lattices that (i) $E_z^{\text{eff}}(h)$ goes to zero precisely at the transition from the KSL to the intermediate phase, i.e.~$E_z^{\text{eff}}(h_c)=0$. (ii) Analyzing the band structure demonstrates that the structure of the Majorana Fermi surfaces obtained at the field-induced transition, $E_z^\text{eff}(h_c)=0$, and in the completely decoupled limit of the corresponding zero-field model, $J_z=0$ and hence $E_z^0=0$, are \textit{identical} (here $E_\gamma^0=\braket{H_\gamma}/N$ indicates energies of the zero-field model). The specific forms of these Fermi surfaces are listed in \tabref{table:summary}. (iii) The nature of the phase for $E_z^\text{eff}(h)>0$, i.e.~the intermediate phase, is identical to that of the phase in the corresponding zero-field model with $E_z^0>0$. Taken together, (i)-(iii) indicate that the physics of the isotropic AFM Kitaev model for $h<h_p$, in particular, the transition into and nature of, the intermediate phase, is intimately related to the physics of the corresponding zero-field model with variable $J_z$ and fixed AFM $J_x$, $J_y$. Note that the sign of the nearest neighbor (NN) spin-spin correlations does not change at the transition, $E_z^\text{eff}(h_c)=0$, but rather within the intermediate phase when $\braket{H_z}=0$. The subsequent transition to the polarized phase at $h=h_p$ is triggered simply when $E_\text{pol}<E_\text{KSL}$. \figref{fig:energies} summarizes this simple picture behind the phase diagram.   

The remainder of the paper is organized as follows: In \secref{sec:MM} we introduce the Kitaev model and explain the fermionization procedure on which the Majorana mean-field theory that is exploited in the following sections is based. \secref{sec:2D} presents the mean-field results for the 2D models, followed by a discussion of the 3D models in \secref{sec:3D}. We end in \secref{sec:Sum} with a discussion and outlook, where we consider potential connections to experiments, limitations of our method and possible extensions thereof.


\section{\label{sec:MM} Model and method}
\subsection{Kitaev model}
The Kitaev honeycomb model \cite{kitaev_anyons_2006} consists of \mbox{$S\!=\!1/2$} spins at the vertices of a honeycomb lattice, coupled via bond-dependent Ising interactions, and is one of the few quantum spin models that can be solved exactly in two dimensions. The Hamiltonian, in the presence of a [001] field, is given in \equref{eq:KitaevHam}.  
In the absence of a magnetic field, two main ingredients are necessary to solve the model. The first is the choice of a clever representation -- or better \textit{fractionalization} scheme -- for the spin degrees of freedom. 
This procedure -- rather routine in the context of frustrated magnets -- often only offers a new starting point for making approximations, e.g. a mean-field decoupling for the emerging four-fermion terms. Fortunately, in the Kitaev model, fractionalization of the spins turns out to be exact. This is due to the fact that we can identify an extensive number of conserved quantities, plaquette \textit{fluxes}, which consitute the second main ingredient. 
Following Kitaev, one can for example substitute each spin degree of freedom by four different flavors of Majorana fermions and then combine them to form a single flavor of itinerant non-interacting Majoranas coupled to a static $\mathbb{Z}_2$ gauge field. After fixing the gauge field sector, the system is then described by its Majorana band structure.
For the honeycomb model, one finds two gapless Majorana cones, at $K$ and $K^\prime$, for approximately equal couplings on all bonds and a gapped phase if one exchange dominates. 
It was quickly realized that generalizations of the model to other lattices, even in three dimensions, inherit the analytical solvability, as long as the lattices remain tricoordinated. As multifaceted as the possible lattices are, so are the resulting Majorana band structures: For isotropic couplings, gapped band structures \cite{yang_mosaic_2007}, both topologically trivial and non-trivial, are possible as are Dirac points and Weyl points, nodal lines and Fermi surfaces in three dimensions \cite{obrien_classification_2016, mandal_exactly_2009, hermanns_quantum_2014, hermanns_weyl_2015}. Note that the ground-state energy and band structure do not depend on the sign of the Kitaev coupling, whether FM or AFM. However, this no longer applies in the presence of a magnetic field.

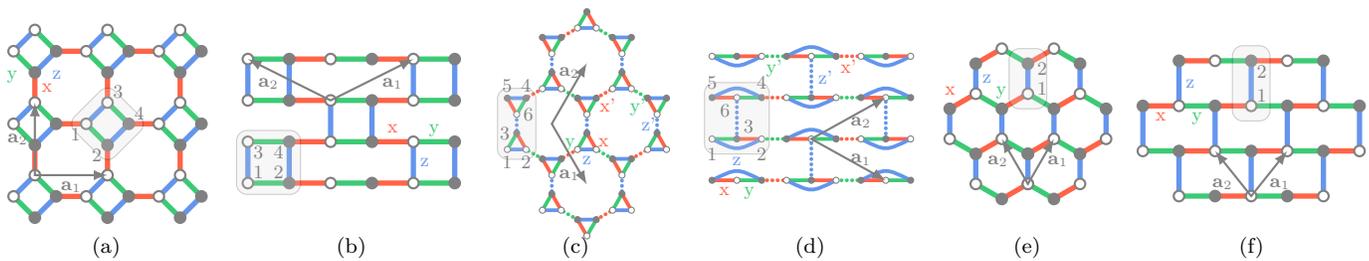
\begin{figure*} 
\begin{tikzpicture} 
\node[inner sep = 0pt]{\includegraphics[scale=1]{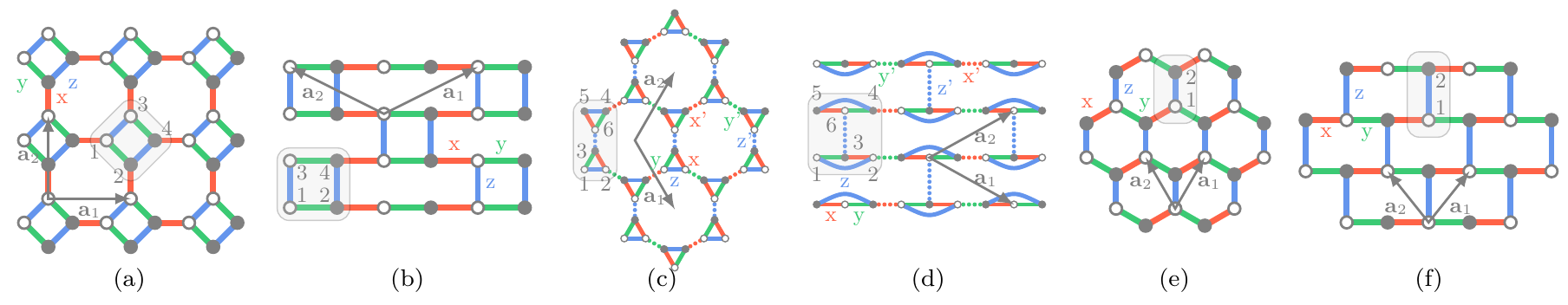}};
\node[opacity=0]{\subfloat[][\label{fig:SO}]{}}; 
\node[opacity=0]{\subfloat[][\label{fig:SObrick}]{}}; 
\node[opacity=0]{\subfloat[][\label{fig:YK}]{}};
\node[opacity=0]{\subfloat[][\label{fig:YKbrick}]{}};
\node[opacity=0]{\subfloat[][\label{fig:HC}]{}}; 
\node[opacity=0]{\subfloat[][\label{fig:HCbrick}]{}}; 
\end{tikzpicture}
\caption{\textbf{Kitaev model, unit cells and translation vectors for 2D lattices:} \protect\subref{fig:SO} square-octagon, \protect\subref{fig:YK} decorated honeycomb lattice and \protect\subref{fig:HC} honeycomb. The topologically equivalent brick lattice versions are shown in \protect\subref{fig:SObrick}, \protect\subref{fig:YKbrick} and \protect\subref{fig:HCbrick}. The Jordan-Wigner strings correspond to the $xy$-chains running from left to right in the brick lattice versions.}
\label{fig:2Dlattices}
\end{figure*}

\subsection{Jordan-Wigner transformation}
The Kitaev model can be fermionized using a Jordan-Wigner transformation \cite{mandal_rvb_2012,feng_topological_2007,chen_exact_2008,chen_exact_2007,nasu_vaporization_2014}. 
Compared to Kitaev's original solution \cite{kitaev_anyons_2006} or alternative fractionalization schemes \cite{burnell_su2_2011}, it has the advantage that the Hilbert space is not enlarged and we are not faced with the technical difficulty of imposing constraints to obtain physical states. 
At the same time, the transformation is ideally suited to coupling the spins to a [001] field due to the simple form of $S^z$. 

In order to apply the transformation, the lattices are considered as arrangements of chains consisting of $x$- and $y$-bonds that are connected by $z$-bonds. For each chain, the Jordan-Wigner transformation
\begin{align}
	S^+_{mn} & = \frac{1}{2} a_{mn} ^\dagger e^{i\pi \sum \limits_{l<n} a_{ml}^\dagger a_{ml}},  	\label{jordanwigner1} \\
	S^z_{mn} & = \frac{1}{2} \left( 2a_{mn}^\dagger a_{mn} - 1 \right) \label{jordanwigner2},
\end{align}
is applied.
The index $m$ labels the chain and the index $n$ the position within the chain. For the 2D lattices, $m$ ($n$) corresponds to the row (column) number of the topologically equivalent brick lattices, see \figref{fig:2Dlattices}. For the 3D lattices see \figref{3Dlattices}. Each lattice is divided in two sublattices, colored in black ($b$) and white ($w$), such that sites belonging to different sublattices alternate on each chain. The $z$-bonds can either connect sites of different color (e.g. honeycomb lattice), same color (e.g. square-octagon) or both (e.g. decorated honeycomb). Using \equref{jordanwigner1} and \equref{jordanwigner2}, the interactions are rewritten as
\begin{align}
	S^x_{mn} S^x_{mn+1}  =& - \frac{1}{4} \left( a_{mn}-a_{mn}^\dagger \right) \left(a_{mn+1} + a_{mn+1}^\dagger \right), \label{jordanwigner_xx} \\
	S^y_{mn} S^y_{mn+1}  =&  \frac{1}{4} \left( a_{mn}+a_{mn}^\dagger \right) \left(a_{mn+1} - a_{mn+1}^\dagger \right), \label{jordanwigner_yy} \\
	S^z_{mn} S^z_{kl}   =& \frac{1}{4}  \left( 2a_{mn}^\dagger a_{mn} - 1 \right)  \left( 2a_{kl}^\dagger a_{kl} - 1 \right).
\end{align}
We introduce Majorana operators $c$ and $\bar{c}$ on the two sublattices according to
\begin{align}
	c_w &=  a+a^\dagger, & \bar{c}_w &= -i \left(  a-a^\dagger \right), \\
	\bar{c}_b &=  a+a^\dagger, & c_b &= -i \left(  a-a^\dagger \right).
\end{align}
The Kitaev Hamiltonian then becomes 
\begin{align}
	H =  &\frac{iJ_x}{4} \sum \limits_{x \text{-bonds}} c_b c_w -\frac{i J_y}{4}  \sum \limits_{y \text{-bonds}}  c_w c_b \nonumber \\
	&-\frac{i J_z}{4} \sum \limits_{z \text{-bonds}} \delta_r \alpha_r c_i c_j  \nonumber \\
	&- \frac{i \h}{2} \left( \sum \limits_b  c_b\bar{c}_b - \sum \limits_w c_w\bar{c}_w \right), \label{eq:JW-Hamiltonian}
\end{align}
where $i$ and $j$ in the third sum are connected via the $z$-bond $r$ and in the last sum, $b$ ($w$) numbers all black (white) lattice sites and $\delta_r$ equals $-1$ if the $z$-bond $r$ connects two sites of the same color and $1$ otherwise. So far, \equref{eq:JW-Hamiltonian} is an exact rewriting of \equref{eq:KitaevHam}. For $\h=0$, $ \alpha_r = i \bar{c}_i \bar{c}_j$ is a conserved $\mathbb{Z}_2$ quantity defined on each $z$-bond. It effectively plays the role of the gauge field in the original solution, where these $\mathbb{Z}_2$ conserved quantities are defined for all bonds.

The conserved loop or plaquette operator to whom the Kitaev model owes its exact solvability is defined as
\begin{equation}
W_p = \prod \limits_{(i,j)_\gamma \in \partial p} S_i^\gamma S_j^\gamma \,,
\label{loop_operator}
\end{equation}
where the product is taken over all bonds that form the plaquette $p$ \footnote{This form of $W_p$ is consistent with Lieb's definition of a flux \cite{lieb_flux_1994}. In an alternative definition, the product in \equref{loop_operator} comprises only the spin component not included along the loop. 
	Depending on the lattice, these two definitions can differ by a sign, e.g. for the square-octagon lattice.}: 
Its eigenvalue determines the $\mathbb{Z}_2$ flux through the plaquette. An eigenvalue of $+ 1 (-1)$ corresponds to $0 (\pi)$-flux. In the original solution, this eigenvalue is given by the product of the gauge variable on every bond. As the Jordan-Wigner transformation corresponds to a {\em gauge fixing} procedure for all bonds lying in the Jordan-Wigner chains \cite{mandal_rvb_2012}, only $z$-bonds remain in the product and we get
\begin{equation}
W_p = \prod \limits_{(i,j)_z \in \partial p}  \bar{c}_i \bar{c}_j \,.
\label{loop_operator2}
\end{equation}	
Note that time reversal symmetry is broken in the flux sector if this product contains an {\em odd} number of the operators $\bar{c}_i \bar{c}_j$ with purely imaginary expectation value.

\subsection{Mean-field theory}
For $\h=0$, replacing $\alpha_r$ by $\pm1$ results in a free hopping problem for the $c$-Majorana fermions and we recover Kitaev's exact solution.
The exact solubility is no longer given when the field is switched on due to the hybridization between $c$- and $\bar{c}$-Majoranas that is generated by the last term in \equref{eq:JW-Hamiltonian}. In the Jordan-Wigner language, the system is composed of interacting Majoranas along the $z$-bonds and freely hopping Majoranas on $x$- and $y$-bonds. The interactions are decoupled via
\begin{align}
	i \bar{c}_i \bar{c}_j i c_i c_j  \approx & i \bar{A}_{ij} c_i c_j  + iA_{ij} \bar{c}_i \bar{c}_j - \bar{A}_{ij} A_{ij} +i \bar{B}_{ij}\bar{c}_j c_i  \nonumber \\   + & iB_{ij} \bar{c}_i c_j - \bar{B}_{ij} B_{ij} - i \Gamma_i \bar{c}_j c_j  - i \Gamma_j \bar{c}_i c_i +  \Gamma_i \Gamma_j, 
\end{align}
where the real mean-field parameters 
\begin{align}
	A_{ij} = \langle i c_i c_j \rangle, && \bar{A}_{ij} =  \langle 	i \bar{c}_i \bar{c}_j \rangle, \nonumber \\
	B_{ij} = \langle i \bar{c}_i c_j \rangle, && \bar{B}_{ij} = \langle i \bar{c}_j c_i \rangle, && \Gamma_l = \langle i \bar{c}_l c_l \rangle,
\end{align}
were introduced. For the white (black) sublattice, the quantity $\Gamma_l$ corresponds to twice the (negative) magnetization in $z$-direction $\langle S^z_i \rangle \equiv M^z_i $.
Assuming translational invariance, the Fourier transformed Hamiltonian
\begin{equation}
H^{\text{MF}} = \sum \limits_{\vect{k}} c^\dagger_\vect{k} H(\vect{k}) c_\vect{k}
\label{eq:kspace-Hamiltonian}
\end{equation} 
is used to solve the mean-field equations self-consistently. The sum in \equref{eq:kspace-Hamiltonian} is restricted to half of the Brillouin zone to avoid double counting of states and in $c_\vect{k} = \left( c_{\vect{k},1},\dots, c_{\vect{k},N}, \bar{c}_{\vect{k},1}, \dots, \bar{c}_{\vect{k},N} \right)^\text{T}$, the index labels the sites within a unit cell.


\section{Results}
We now devote ourselves to a detailed discussion of our results on the Kitaev model in the presence of a uniform magnetic field in the [001] direction, summarized in \tabref{table:summary}, \figref{fig:genericPD} and \figref{fig:energies}. In the following, we use the terminus `polarized phase' for the phase that is adiabatically connected to the fully polarized state with $M^z\! =\! 1/2$ at $h \to \infty$. The phase that smoothly evolves from the KSL at zero field is referred to as the KSL (or $B$, $A_i$), even at finite $\h$. 
Unless otherwise noted, we consider isotropic couplings and set $|J|\!=\!1$.  
Detailed information about the lattices can be found in \appref{appendix:Hamiltonian}, together with explicit expressions of the original spin Hamiltonian and their Jordan-Wigner forms.
The discussion of the mean-field results is supplemented by \appref{appendix:MoreResults}, where more details on form and evolution of the mean-field parameters and properties of the Majorana band structure are provided.

\subsection{\label{sec:2D}2D Kitaev models}	
We start the detailed discussion of the results with the two-dimensional lattices that are shown in \figref{fig:2Dlattices}. 
For all three lattices, we follow the same scheme and discuss first the FM Kitaev model and then the AFM one. Finally, we consider the influence of anisotropic couplings in the AFM model to shed more light on the nature of the intermediate phases appearing in the isotropic case.


\subsubsection{\label{subsec:SO} Square-octagon lattice}
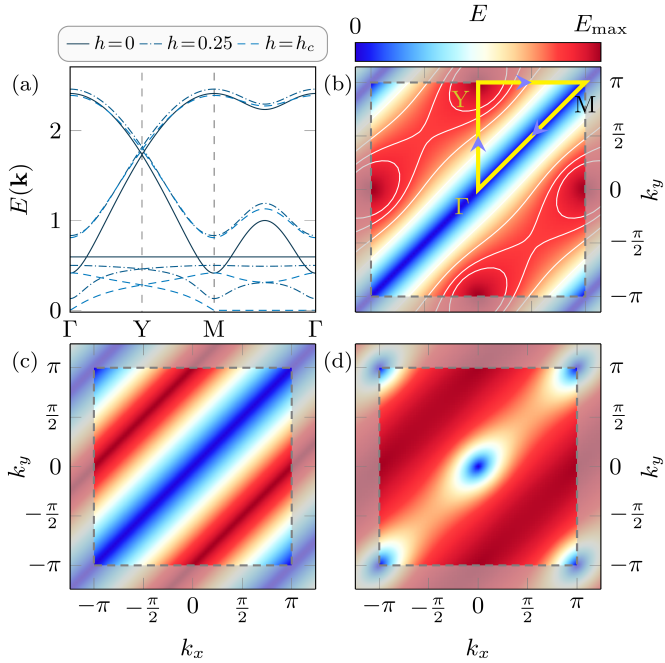
\begin{figure}
	\hspace*{-0.4cm}
	\centering
	\begin{tikzpicture} 
	\node at (0,0){\includegraphics{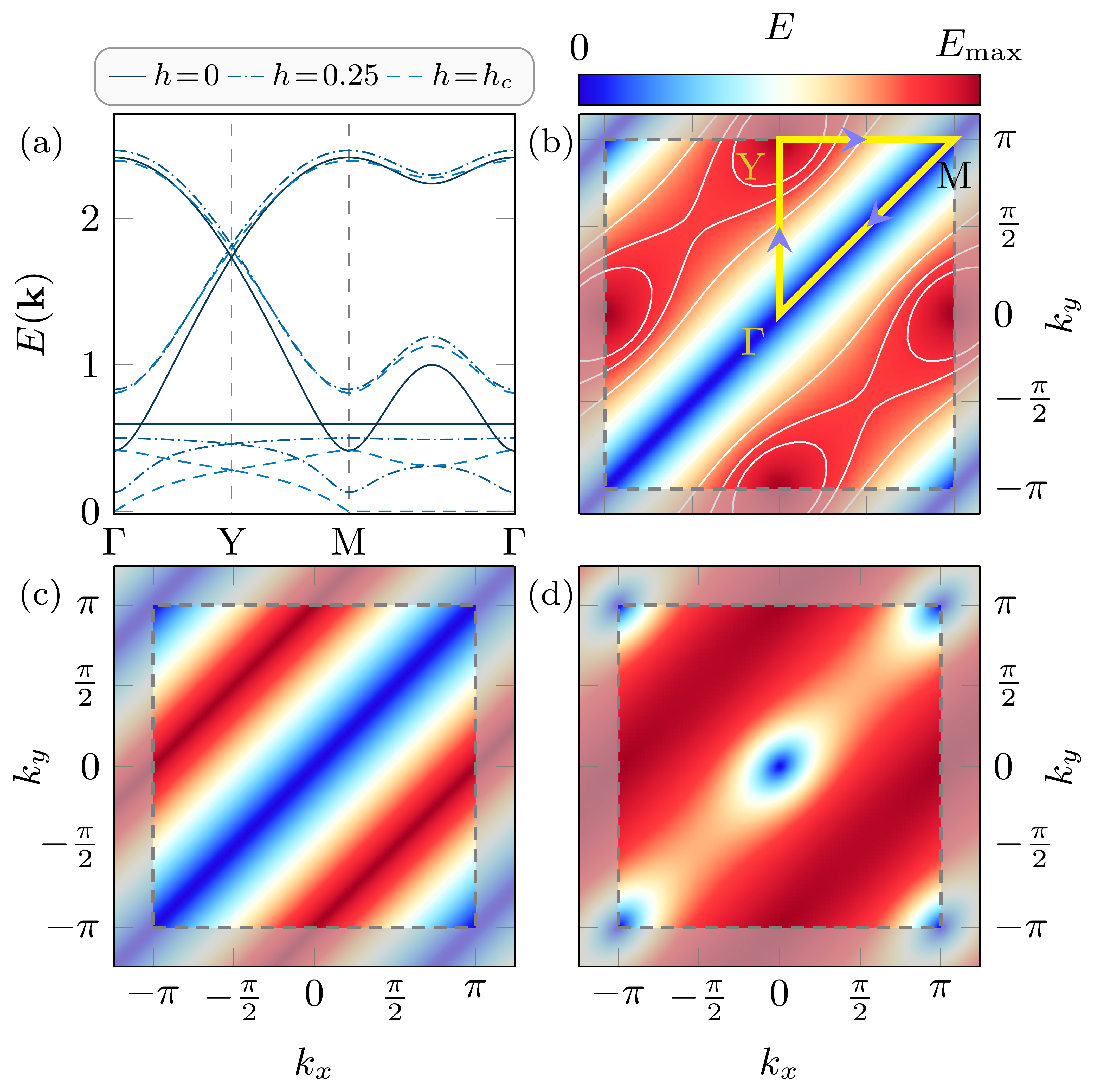}}; 
	\node[opacity=0]{\subfloat[][\label{fig:SO_1D_bands}]{}}; \node[opacity=0]{\subfloat[][\label{fig:SO_2D_bands_hc1}]{}};
	\node[opacity=0]{\subfloat[][\label{fig:SO_2D_bands_Jz0}]{}}; \node[opacity=0]{\subfloat[][\label{fig:SO_2D_bands_B}]{}};
	\end{tikzpicture}
	\caption{\textbf{Mean field band structure for the square-octagon lattice.} \protect \subref{fig:SO_1D_bands} Cut along the high symmetry lines that are shown in \protect \subref{fig:SO_2D_bands_hc1}. The flat bands for $\h=0$ correspond to $\bar{c}$-Majoranas that are localized on $z$-bonds. \protect \subref{fig:SO_2D_bands_hc1} Energy of the lowest band for isotropic couplings and $h_{c}$ with a gapless mode along the diagonal of the first Brillouin zone (marked by the dashed line). \protect\subref{fig:SO_2D_bands_Jz0} Energy of the lowest band for $J_z=0$ and $J_x=J_y$. \protect\subref{fig:SO_2D_bands_B} Energy of the lowest lying band for the $B$ phase at $h\!=\!0.075$, $J_x\!=\!-0.38$ and $J_y\!=\!J_z\!=\!-0.31$.}
\end{figure}

\begin{figure} 
\hspace{-2.1cm}
	\begin{tikzpicture}
		\node[inner sep = 0pt]{\includegraphics[trim = 0cm 0cm 0cm 0cm, clip=true]{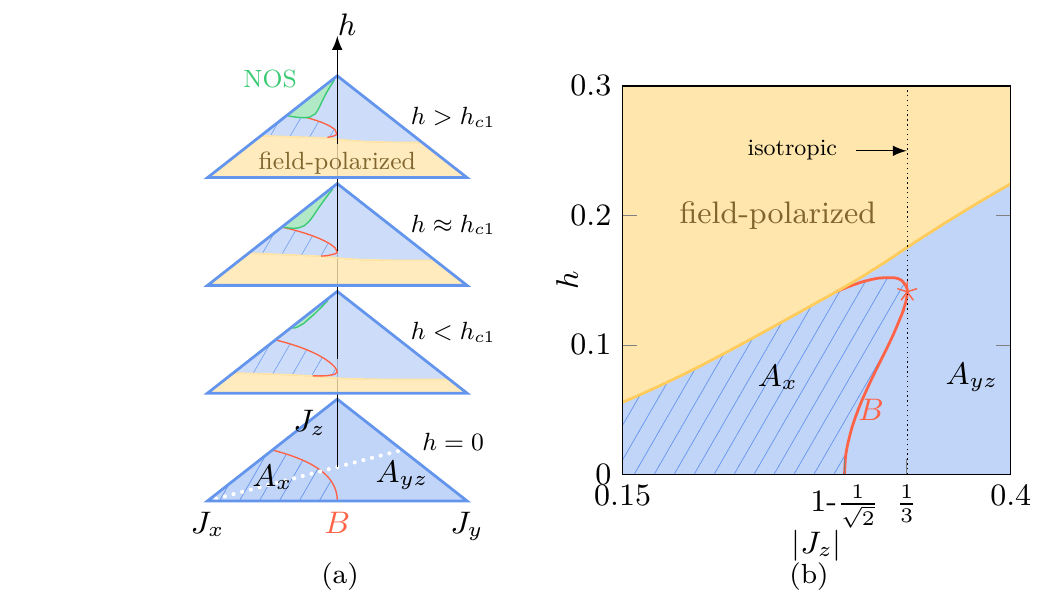}};
		\node[opacity=0]{\subfloat[][\label{fig:SO_AFM_aniso1}]{}}; \node[opacity=0]{\subfloat[][\label{fig:SO_AFM_aniso2}]{}};
	\end{tikzpicture} 
	\caption{\protect\subref{fig:SO_AFM_aniso1} \textbf{Phase diagram of the Kitaev model on the square-octagon lattice in a [001] field with anisotropic AFM couplings.} With increasing field strenght, the polarized phase covers more and more space, the border between $A_x$ and $A_{yz}$ deforms and a new phase (NOS), characterized by spins aligned antiparallel along $z$-bonds emerges for $|J_x|,|J_z|\!>>\!|J_y|$. \protect\subref{fig:SO_AFM_aniso2} Phase diagram for $\h$ and the parameters chosen along the line defined by $J_y=J_z$ and $|J_x|\!+\!|J_y|\!+\!|J_z|\!=\!1$ (white dotted line in \protect\subref{fig:SO_AFM_aniso1}). The point where the gap closes along a line in $k$-space is denoted by the red star. From this perspective, it appears that the first phase transition is just a touching of the $B$ phase.}
\end{figure}

The Kitaev model on the square-octagon lattice was first studied by \citet{yang_mosaic_2007} using Kitaev's original approach.
The lattice and its deformed brick-wall version are shown in \figref{fig:SO} and \figref{fig:SObrick}.
We start the discussion by briefly reviewing the results for $\h=0$ using Jordan-Wigner fermions and relate them to \refref{yang_mosaic_2007}. Considering a four-site unit cell as depicted in \figref{fig:SO}, we find two different mean-field solutions for all coupling configurations, corresponding to the flux-free sector, with zero flux through every plaquette, and the full-flux sector, with a flux of $\pi$ through both the square and octagon plaquettes. The ground state is found to belong to the full-flux sector, in accordance with Lieb's theorem \cite{lieb_flux_1994} for lattices with loop length $0 \!\mod 4$ \footnote{Note that a different definition of the loop operator is used in \refref{yang_mosaic_2007}, which, in that language, results in a flux-free ground state \cite{hermanns_quantum_2014}.}.
The condition for the Majorana band structure to be gapless is $J_z^2=J_x^2-J_y^2$, leading to the phase diagram shown at the bottom of \figref{fig:SO_AFM_aniso1}. 
The spectrum for $k_x=0$ and isotropic coupling constants is shown in \figref{fig:SO_1D_bands} \footnote{Compared to \figref{fig:SO_1D_bands}, the band structure in \refref{yang_mosaic_2007} is shifted by $(\pi,0)$, the reason being that the Jordan-Wigner transformation induces a different gauge structure for $x/y$-bonds \cite{mandal_rvb_2012}. In \refref{yang_mosaic_2007}, choosing a different gauge that still ensures the full-flux sector restores the band structure shown here.}.

For AFM (FM) couplings, the exact solution for the ground-state energy $E_0=-0.804$ per unit cell is reproduced with the non-vanishing mean-field parameters 
\begin{align}
	A_{13} &= -A_{24} = \pm (\mp) 0.595,  \nonumber \\
	\bar{A}_{13} &= -\bar{A}_{24} = \pm 1.  \label{SO_mf_2}
\end{align}
The latter ensures that  
\begin{equation}
\langle W_\square \rangle=\langle \bar{c}_1 \bar{c}_3 \bar{c}_4 \bar{c}_2 \rangle = \frac{1}{i^2} 	\bar{A}_{13} 	\bar{A}_{42} =  -1 
\label{eq:WSO}
\end{equation}
for all square plaquette operators and similarly for the octagon plaquettes $W_{\octagon{0.5}}$. For $\h>0$, the mean-field parameters for both AFM and FM couplings fulfill the relations
\begin{align}
	A_{13} &= -A_{24}, \nonumber \\ 
	\bar{A}_{13} &= -\bar{A}_{24}, \nonumber \\
	B_{13} &= B_{24} =	\bar{B}_{13} = \bar{B}_{24} = 0,  \nonumber \\
	M^z_1 &= M^z_2 = M^z_3 = M^z_4.
\end{align}

\paragraph{FM couplings}


\begin{figure} 
\begin{tikzpicture} 
\node[inner sep =0pt]{\includegraphics[scale=1]{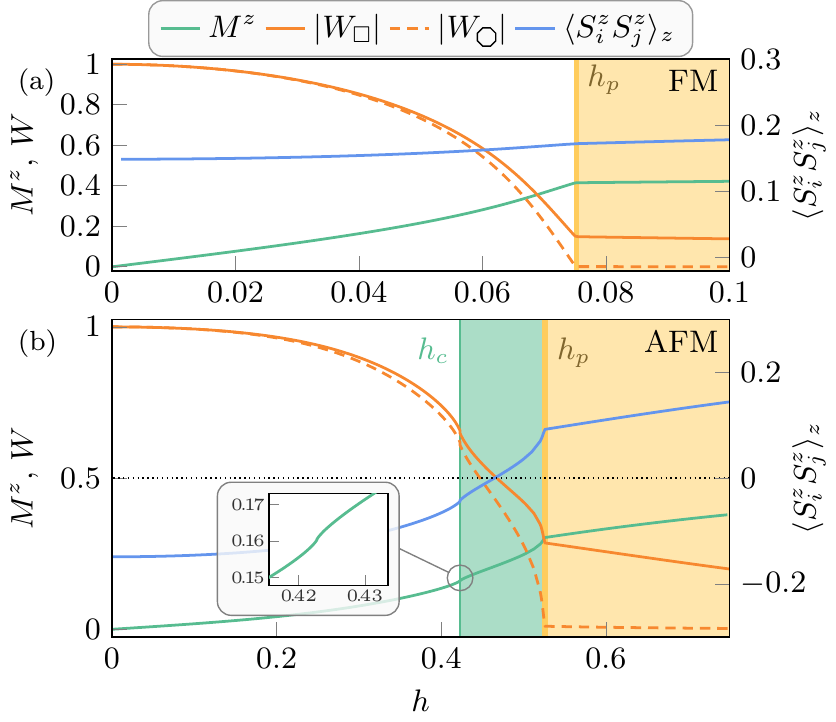}};
\node[opacity=0]{\subfloat[][\label{fig:SO_FM}]{}};\node[opacity=0]{\subfloat[][\label{fig:SO_AFM}]{}};
\end{tikzpicture} 
\caption{\textbf{Summary of the results for the Kitaev model on the square-octagon lattice under a magnetic field in [001] direction} for \protect\subref{fig:SO_FM} FM and \protect\subref{fig:SO_AFM} AFM couplings. Shown here are the magnetization, the expectation value of the flux operator for the two different plaquettes and the NN correlations $\langle S^z_iS^z_j \rangle_z$ as a function of $\h$.}
\label{fig:SOsummary}
\end{figure}

\figref{fig:SO_FM} displays the main results for FM couplings. 
The magnetization increases rapidly with increasing field strength and a single phase transition to the field-polarized phase occurs at a critical field strength of $h_c\! = \! 0.075$. This is the highest critical field for all FM models considered in this study. The band structure is gapped for all $\h \geq 0$. In the polarized phase, all mean-field parameters disappear, apart from the magnetization. However, there is a residual flux through the square plaquettes that approaches zero only in the limit $\h \to \infty$. More details on mean-field parameters, correlation functions and susceptibility are presented in \appref{appendix:MoreResults}.

\paragraph{AFM couplings}
The results from the self-consistent calculations for AFM couplings are summarized in \figref{fig:SO_AFM}.
There are two phase transitions present.
A kink in the evolution of the fluxes and magnetization indicates a second-order transition at $h_{c}=0.423$.
Then, at $h_{p}=0.527$, all mean-field parameters, except for the magnetization, vanish and the polarized phase appears via a continuous transition.

As noted in \secref{sec:Overview}, by defining $E_z^\text{eff}(h)$, \equref{eq:Ezeff}, we see that the first phase transition, at $h_c$, is simply due to a sign change in $E_z^{\text{eff}}(h)$. 
Furthermore, the effect of, on the one hand, the magnetic field and, on the other hand, of anisotropies in the coupling constants in the absence of a field, on the low-energy spectrum are the same. For isotropic couplings the zero-field band structure is shown in \figref{fig:SO_1D_bands}, with the flat bands arising from $\bar{c}$-fermions localized on the $z$-bonds. On increasing $\h$, the two different Majorana flavors hybridize and all bands become dispersive. The energy gap closes quadratically, giving rise to a nodal line at the transition $E_z^\text{eff}(h_c)=0$, as illustrated in \figref{fig:SO_2D_bands_hc1}. Note that -- as the white contour lines indicate -- only the nodal line is dispersionless. The low-energy excitations are governed by both, $c$- and $\bar{c}$-Majorana fermions, see also \figref{fig:103b-bands}, whereas the $\bar{c}$-sector is not present for $h\!=\!0$. On the other hand, in the absence of a field, by setting $J_z =0$ and hence $E_z^0=0$, the lattice decouples into chains and the spectrum is again dispersionless with a nodal line along the diagonal of the Brillouin zone, as shown in \figref{fig:SO_2D_bands_Jz0}, the same nodal line that appears for $E_z^\text{eff}(h_c)=0$. Indeed, the condition for gapless modes to appear in the zero-field model, $J_z^2=J_x^2-J_y^2$, is simply replaced by $(J_z E_z^\text{eff} / E_z^0)^2 =J_x^2-J_y^2$.
Note that the NN spin-spin correlations along $z$-bonds change from AFM to FM at a field strength just above $h_{c}$, see \figref{fig:SO_AFM}.

As shown in \figref{fig:SO_AFM}, the fluxes $\langle W \rangle$ through the square and octagon plaquettes exhibit a similar evolution in the KSL phase. Between $h_c$ and $h_p$, $\langle W_{\octagon{0.5}}\rangle$ drops rapidly, nearly approaching zero at $h_{p}$, whereas $\langle W_\square \rangle$ decreases slower, having a large residual value even in the field-polarized phase.

\paragraph{Anisotropic couplings} 


We can gain more insight into the nature of the transition at $h_{c}$ by allowing for anisotropic couplings.
To do so, we consider varying the AFM coupling strengths, while keeping $\sum_i |J_i| = 1$. As depicted in figure \figref{fig:SO_AFM_aniso1}, the phase diagram at $\h=0$ consists of two gapped phases, $A_x$ and $A_{yz}$, named after the dominating coupling constants $J_x$ and $J_{y/z}$ respectively \cite{yang_mosaic_2007}. The transition between the two phases is accompanied by the emergence of Dirac cones at $\Gamma$ and $M$. This line separating $A_x$ and $A_{yz}$ is denoted as $B$. As for the honeycomb lattice, a magnetic field in the $[111]$ direction induces an effective hopping between next nearest neighbors and the $B$ line expands into a gapped phase covering a finite region in parameter space \cite{yang_mosaic_2007, kells_kaleidoscope_2011}. The Chern number of the now $B$ phase is $\nu = \pm 1$ at low fields, with anisotropies in the couplings facilitating phases with even higher Chern number, up to $\nu = \pm 4$ \cite{kells_kaleidoscope_2011}. 

The phase diagram for the model under a $[001]$ field is presented in \figref{fig:SO_AFM_aniso1}. In contrast to the findings for a field in the [111] direction, the $B$ line does not spread out but remains a line, though it does deform and shift within parameter space. The gapless points along the $B$ line always appear at $\Gamma$ and $M$, as shown in \figref{fig:SO_2D_bands_B} for $\h\!=\!0.075$ and $J_y\!=\!J_z \!=\! -0.31$. The KSL is more stable for larger AFM $J_z$ and, as the field increases, the polarized phase covers an increasingly large region in the phase diagram, starting from the lower edge where $J_z=0$.

For $J_x\!=\!J_y\!=\!0$, the lattice decouples into isolated dimers and the KSL solution simply corresponds to isolated dimer singlets and is degenerate in energy with a solution in which spins align antiparallel along $z$-bonds with $|M_i^z|=1/2$. 
In the latter solution, spins connected by $y$-bonds can point either in the same or opposite direction, resulting in stripy or N\'eel order on each square.
A finite $J_x (J_y)$ prefers an arrangement smoothly connected to the N\'eel (stripy) order. If $J_{x,y},\h\!\neq \!0$, these states can lower their energy by suppressing the magnetization pointing opposite to the field direction. Indeed, for $|J_x|\!>\!|J_y|$, the phase descending from the N\'eel-ordered squares (NOS) actually becomes preferable over the KSL, occupying a finite region of parameter space, as shown in \figref{fig:SO_AFM_aniso1}. However, for $|J_y|\!>\!|J_x|$, the phase evolving from the stripy-ordered squares never becomes the groundstate.

How can these observations enlighten our understanding of the transition at $h_c$ in the case of isotropic couplings? \figref{fig:SO_AFM_aniso2} shows the phase diagram along the white line indicated at the bottom of \figref{fig:SO_AFM_aniso1}. For increasing field strength, the $B$ line moves to lower values of $J_z$ and the lowest energy band between the $M$ and $\Gamma$ points becomes increasingly flat. At $h_c$, the $B$ line reaches the isotropic point and there are zero-energy modes for all $\textbf{k}$ between $M$ and $\Gamma$. The transition is thus simply a touching of the critical $B$ line. Both sides of $h_{c}$ lie within the $A_{yz}$ phase, and are smoothly connected once anisotropy is introduced. Note that starting slightly to the left of the isotropic cut in \figref{fig:SO_AFM_aniso2} enables a sequence of field-induced transitions from $A_{yz} \to A_x \to A_{yz} \to \text{polarized}$.


\subsubsection{Decorated honeycomb lattice}		
Inflating each site of the honeycomb lattice to a triangle, whose size is chosen such that all bonds have the same length, generates a new lattice which we refer to here as the decorated honeycomb lattice (also known as the triangle-honeycomb or Fisher lattice). 
As already foreseen by Kitaev \cite{kitaev_anyons_2006}, and fully elucidated by Yao and Kivelson \cite{yao_exact_2007}, new physics arises due to the {\em odd} number of bonds forming the triangular plaquettes: 
The eigenvalue of $W_\vartriangle$ is $\pm i$ and therefore odd under time reversal symmetry (TRS) implying that TRS is spontaneously broken upon the transition into the QSL regime. Such a QSL with broken TRS is called a \textit{chiral spin liquid} (CSL) \cite{kalmeyer_equivalence_1987}.

As shown in \figref{fig:YK}, two sets of coupling parameters not connected by symmetries can be distinguished, $J_i$ on the bonds forming the up and down pointing triangles and $J'_i$ on the bonds connecting them.
For isotropic $J$  and $J^\prime$, two different ground states are possible depending on the ratio $J'/J$ \cite{yao_exact_2007}: A topologically trivial CSL for $J'/J\!>\!\sqrt{3}$ and a non-trivial CSL with Chern number $\nu=\pm1$ otherwise. The non-trivial CSL hosts anyonic excitations with non-Abelian braiding statistics \cite{kitaev_anyons_2006, yao_exact_2007, dusuel_perturbative_2008}. 
For $\h=0$ and AFM couplings $J\!=\!J'\!=\!-1$, the mean-field solution with the lowest energy and $\nu=+1$ is given by
\begin{align}
	A_{12} &= A_{45} = \pm 0.482,  && A_{36} = \pm 0.566, \nonumber \\
	\bar{A}_{12} &= \bar{A}_{45} = \pm 1, && \bar{A}_{36} = \mp 1,  \label{YK_mf_1}
\end{align}
corresponding to the uniform flux configuration that was identified as the ground state flux sector both numerically and -- in the limits $J\!\ll\!J'$ and $J\!\gg\!J'$ -- analytically \cite{yao_exact_2007}. The TRS partner with $\nu\!=\!-1$ is obtained by changing the sign of all mean-field parameters on solid bonds.
For FM couplings, the relative sign between $A$ and $\bar{A}$ is reversed.
We now turn to a discussion of the model in a [001] field, starting again with isotropic FM couplings. 

\begin{figure} 
	\begin{tikzpicture} 
		\node[inner sep =0pt]{\includegraphics{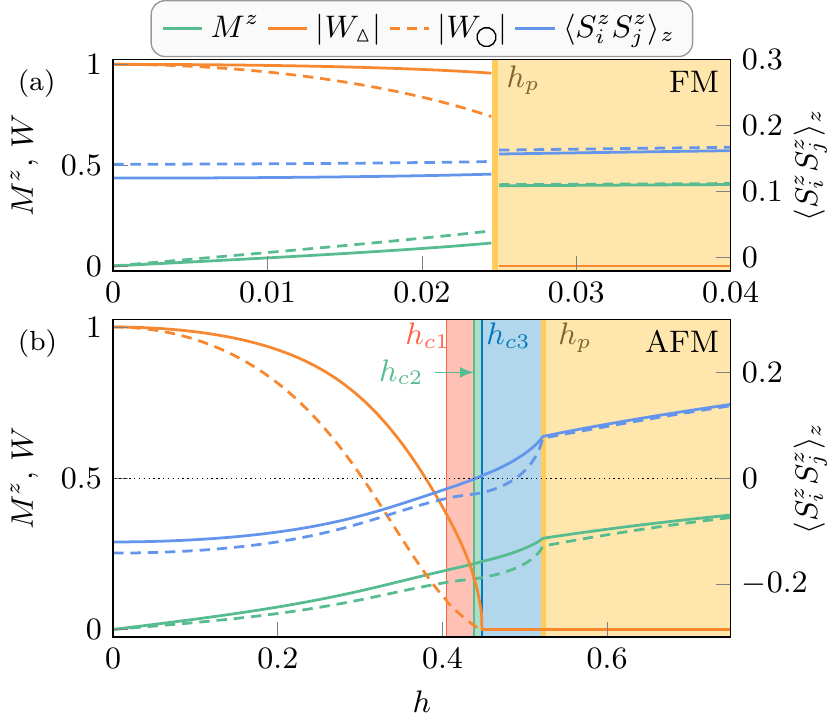}};
		\node[opacity=0]{\subfloat[][\label{fig:YK_FM}]{}}; \node[opacity=0]{\subfloat[][\label{fig:YK_AFM}]{}};
	\end{tikzpicture}
	\caption{\textbf{Summary of the results for the Kitaev model on the decorated honeycomb lattice in a [001] field.} Shown are the evolution of the magnetization, the expectation values of plaquette operators and the NN correlations $\langle S^z_iS^z_j \rangle_z$ for \protect\subref{fig:YK_FM} the FM Kitaev model and \protect\subref{fig:YK_AFM} the AFM Kitaev model. Dashed (solid) lines belong to expectation values on dashed (solid) bonds in \figref{fig:YK}. In \protect\subref{fig:YK_AFM}, the Chern number $\nu$ changes its sign at $h_{c2}$ and jumps to zero at $h_{c3}$. An additional gap closing that does not affect $\nu$ appears at $h_{c1}$.}
\label{fig:YKsummary}
\end{figure}

\paragraph{FM couplings}
Turning on the magnetic field, a phase diagram very similar to the square-octagon lattice is obtained. Only a single first-order transition from the CSL to the polarized phase appears at $h_{c} = 0.025$, that manifests itself in a jump in the magnetization in \figref{fig:YK_FM}. The flux expectation values shrink in the CSL phase, with $W_{\protect\dodecagon{0.6}}$ decreasing fastest. They jump to zero as the polarized phase is entered.

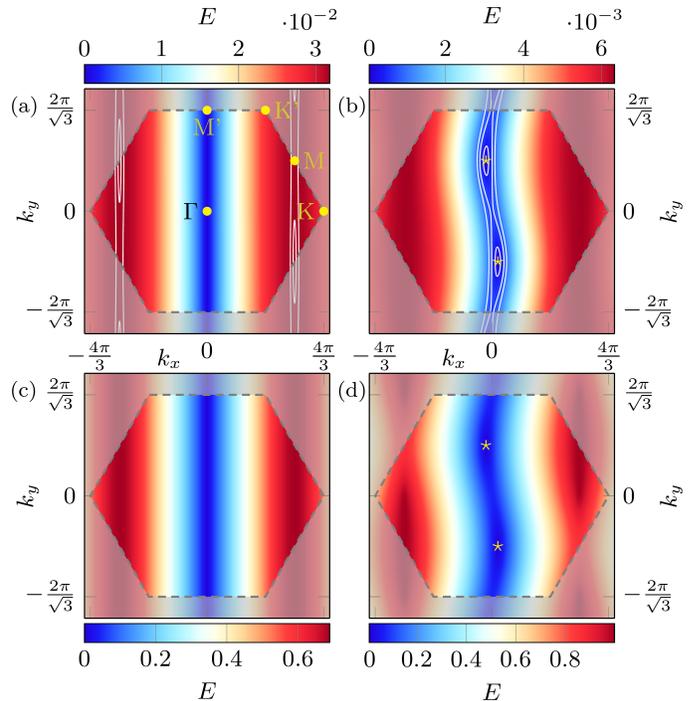
\begin{figure}
	\centering \hspace*{-0.5cm}
	\begin{tikzpicture} 
	\node at (0,0){\includegraphics[trim = 0.0cm 0cm 0cm 0cm, clip = true]{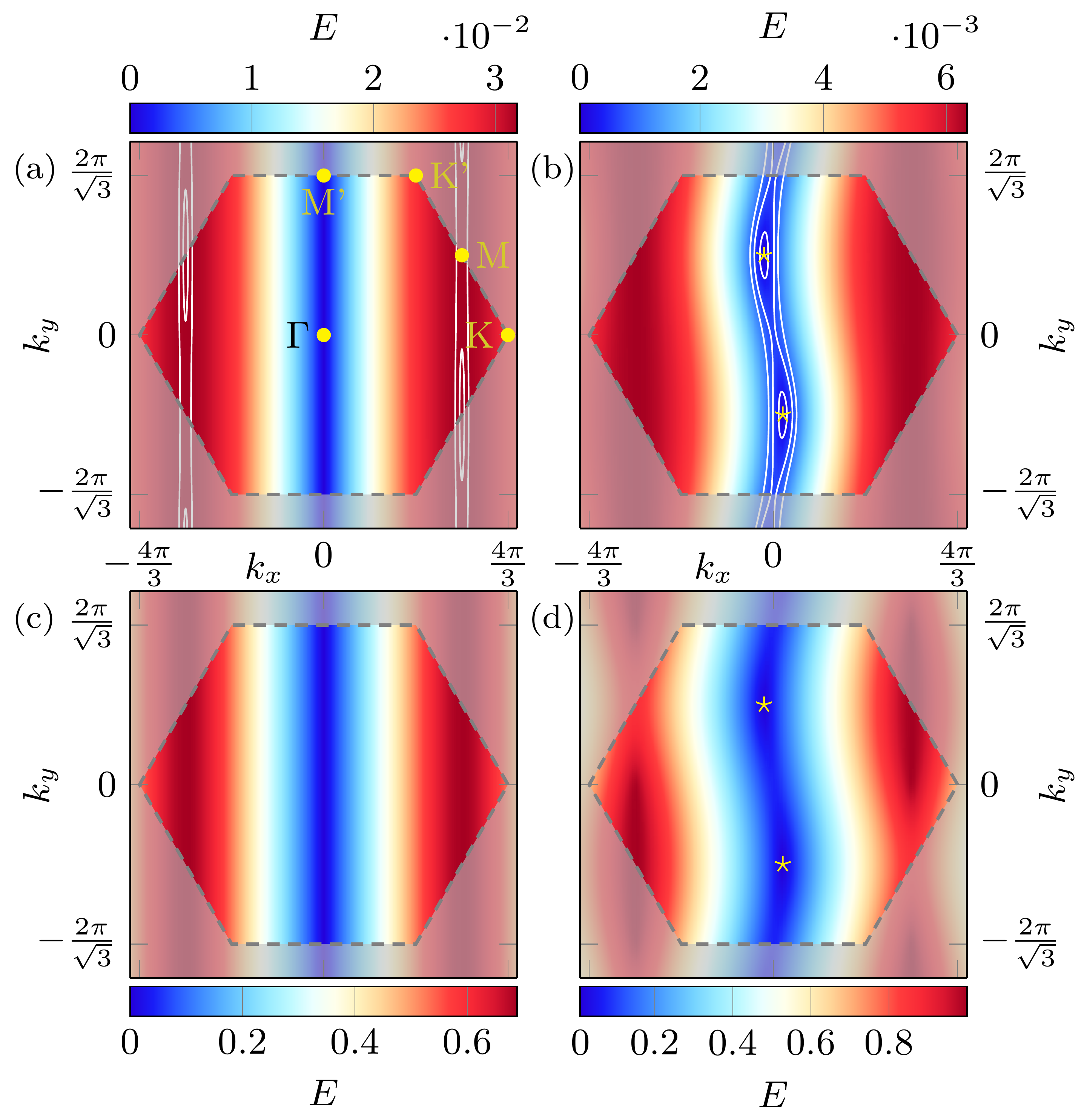}};
	\node[opacity=0]{\subfloat[][\label{fig:YK_2D_bands1}]{}};\node[opacity=0]{\subfloat[][\label{fig:YK_2D_bands2}]{}};
	\node[opacity=0]{\subfloat[][\label{fig:YK_2D_bands3}]{}};\node[opacity=0]{\subfloat[][\label{fig:YK_2D_bands4}]{}};	
	\end{tikzpicture}
	\caption{\textbf{Energy of the lowest band} for isotropic AFM couplings \protect\subref{fig:YK_2D_bands1} at $h_{c1}$, \protect\subref{fig:YK_2D_bands2} at $h_{c2}$ and for $\h=0$ but anisotropic couplings, namely \protect\subref{fig:YK_2D_bands3} $J_z'=0$ and \protect\subref{fig:YK_2D_bands4} $J_z'\approx0.214$ and $J_z=0$. The gapless line that appears at $h_{c1}$ can also be obtained by decoupling the lattice for $J_z'=0$. The gapless points at $h_{c2}$, marked by the yellow stars, also appear for small FM $J_z'$ and $J_z=0$.}
	\label{fig:YK_bands}
\end{figure}

\paragraph{AFM couplings}
For AFM Kitaev interactions, the resulting phase diagram is more complex, as seen in \figref{fig:YK_AFM}.
This can partly be traced back to the existence of different $z$-bonds, as we show below.
The phase diagram exhibits a total of four critical field values, each of which is either characterized by certain mean-field parameters approaching zero or by a gap closing in the excitation spectrum. As for FM couplings, mean-field parameters on different bonds are found to differ with spins on lattice sites 1,2,4 and 5 being stronger polarized.
Starting in a gapped CSL phase with $\nu \! = \! +1$ in zero field, the Majorana gap becomes smaller as $\h$ increases and finally disappears at $h_{c1} = 0.405 $ along the line $k_x=0$ in the first Brillouin zone, as show in \figref{fig:YK_2D_bands1}. Upon further increasing the field, the gap reopens. The Chern number, however, is not affected and remains +1. At $h_{c2} \!=\! 0.438$, the sign of $\nu$ changes, accompanied by a re-closing of the gap at the two isolated points $P \!=\! (-0.214, \pi/\sqrt{3})$ and $-P$. The transition at $h_{c3}\! =\! 0.448$ is characterized by a flat band at $E\!=\!0$. The resulting phase is topologically trivial with $\nu \!=\! 0$. It will be addressed in more detail in the next paragraph. 
Finally, the polarized phase is entered (without a gap closing) via a continuous phase transition at $h_{p} \!=\! 0.523$. The fluxes through both triangular and dodecagonal plaquettes vanish at $h_{c3}$.
Details on the mean-field parameters in each phase are given in \appref{appendix:MoreResults}. 

Again, the nature of the phase transitions at $h_{c1}$ and $h_{c2}$ can be revealed by comparing the low energy spectrum with the spectrum of the zero-field anisotropic model, as done in \figref{fig:YK_bands}.
As a comparison of \figref{fig:YK_2D_bands1} and \figref{fig:YK_2D_bands3} reveals, a dispersionless zero-energy mode is obtained along $k_x=0$ both for $\h=h_{c1}$ with isotropic $J,J^\prime$, and also for $\h=0$ with $J_z^\prime=0$ and all other couplings equal $J_{x,y}^\prime=J_i$. The latter scenario is trivial, as the lattice decouples into one-dimensional chains when $J_z^\prime=0$, resulting in no dispersion along $k_y$. 
The gapless points found at $h_{c2}$, \figref{fig:YK_2D_bands2}, that are shifted slightly away from $k_x=0$, can also be obtained in zero field, but with $J_z=0$ and a small \textit{FM} coupling $J_z'$, for example with the couplings chosen in \figref{fig:YK_2D_bands4}.   

As before, this can be understood by introducing a parameter $E_z^\text{eff}(h)$ whose sign change triggers the field-induced transitions. However, in this case, we need to introduce two such parameters, $E_{z^\prime}^\text{eff}(h)$ and $E_z^\text{eff}(h)$, for the two different $z$-bonds \footnote{Using \equref{eq:Ezeff} to define $E_{z^{(\prime)}}^\text{eff}(h)$, $H_h$ now only contains spins belonging to $z^{(\prime)}$-bonds.}. We find that $E_{z^\prime}^\text{eff}(h_{c1}) =0$ and $E_z^\text{eff}(h_{c2})=0$. The Chern number only changes sign at the second transition, $E_z^\text{eff}(h_{c2})=0$, the same behavior encountered in the zero-field model wherein the Chern number only changes sign when $J_z$ changes sign, and is independent of the sign of $J_z'$. The two transitions can thus be understood as due to a sign change in, first, $E_{z^\prime}^\text{eff}$, and then, $E_z^{\text{eff}}$, mirroring the physics encountered in the zero-field model due to a sign change in, first, $E_{z^\prime}^0$, and then, $E_z^0$. 

\begin{figure}[t]
\includegraphics{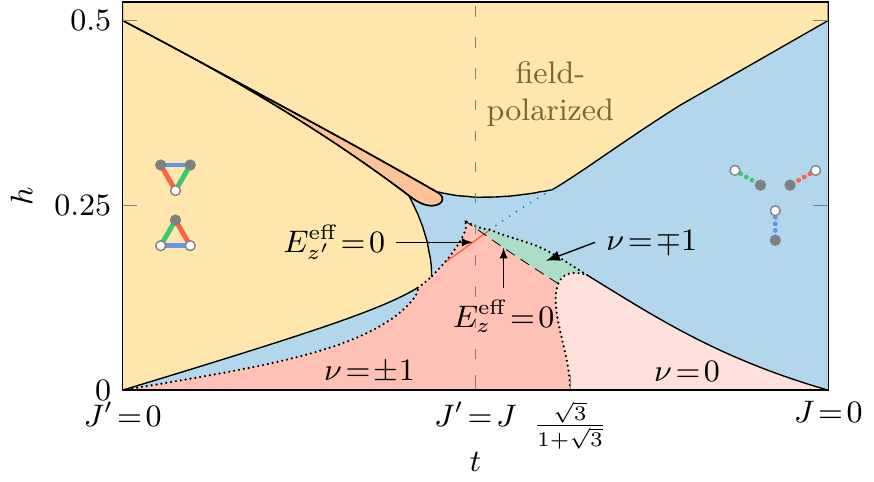}
\caption{\textbf{Phase diagram of the anisotropic AFM Kitaev model on the decorated honeycomb lattice} with coupling constants parametrized as $J'=-t$ and $J=t-1$. Phase transitions where the absolute value of the Chern number changes by 1 are marked by a dotted black line, the black dashed line denotes a sign change in $\nu$.}
\label{fig:YKaniso}
\end{figure}

\paragraph{Anisotropies in $J'/J$}
The phase diagram for anisotropic couplings is illustrated in \figref{fig:YKaniso}. Along the $x$-axis, $J$($J'$) changes linearly from -1 (0) to 0 (-1) with $J' \! + \! J$ being fixed to -1. 
In total we find seven different phases, color-coded in \figref{fig:YKaniso}.
Transitions from topological to trivial phases are marked by dotted lines.
For $\h=0$, we regain the phase diagram from \refref{yao_exact_2007} that shows a phase transition at $J'/J\!=\!\sqrt{3}$ between a topological CSL (red) and a trivial one (light red). 

When either $J$ or $J'$ dominate, intermediate phases extend over a large range of $\h$. If $J\!=\!0,J^\prime\!=\!-1$, the mean-field theory reproduces the exact solution of the decoupled dimer system, sketched on the right of \figref{fig:YKaniso}, which transitions to the polarized phase when $\h\!=\!0.5$. 
The same applies to the opposite scenario, $J'=0,J=-1$, of decoupled triangles. In that case, the intermediate phase and the polarized solution share the same mean-field configuration (and therefore also the same color in the figure).
They are separated by a narrow phase descending from an exact solution at $J'=0$ and $\h=1/2$.
Within the phase sketched in blue, continuously connected to the exact solution for $J\!=\!0,J^\prime\!=\!-1, \h<1/2$, there is a first-order phase transition where the mean-field parameters change discontinuously, marked by the dotted line. All intermediate phases discussed thus far are topologically trivial, and are discussed in more detail in \appref{appendix:MoreResults}.

Let us know turn to the region of the phase diagram where $J \! \approx \! J'$. The topological phase that appeared between $h_{c2}$ and $h_{c3}$ in \figref{fig:YK_AFM} for isotropic couplings is colored in green.
The transition where the sign of $\nu$ changes is indicated by a dashed black line. Along this line, the band structure hosts gapless modes at two isolated points, just as in the zero-field $J_z=0$ case, consistent with $E^{\text{eff}}_z (h)\!= \! 0$ on this line. The same reasoning can be applied to the red line, where the gap closes at $k_x=0$. The appearance of these zero modes is attributed to a sign change in $E^{\text{eff}}_{z^\prime}(h)$ and they are expected to appear at higher fields as $J$ is increased.


\subsubsection{Honeycomb lattice}
\begin{figure}[b]
	\begin{tikzpicture} 
	\node[inner sep = 0pt]{\includegraphics{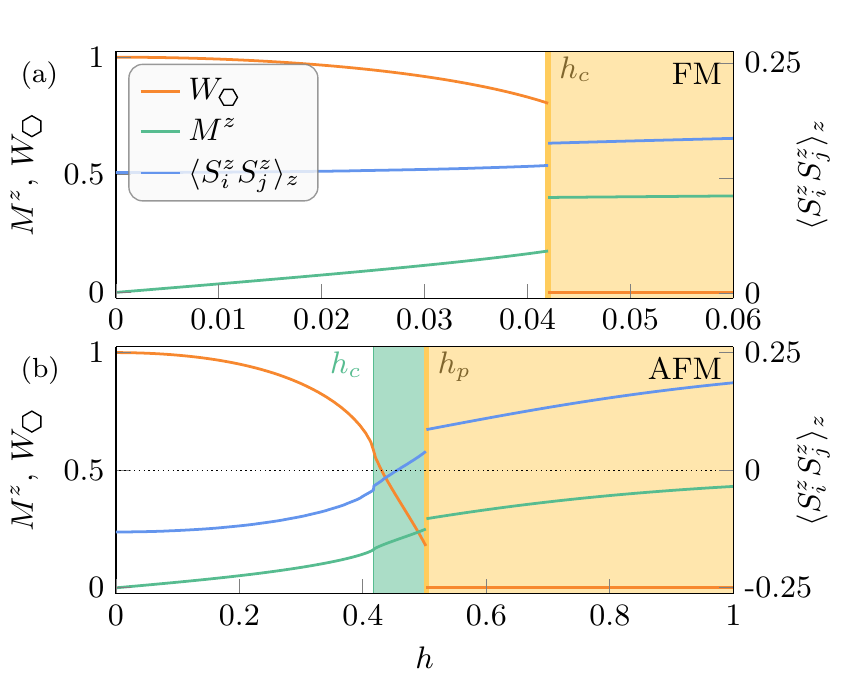}};
	\node[opacity=0]{\subfloat[][\label{fig:HC_FM}]{}};\node[opacity=0]{\subfloat[][\label{fig:HC_AFM}]{}};
	\end{tikzpicture} 
	\caption{\textbf{Summary of the results for the Kitaev model on the honeycomb lattice under a magnetic field in [001] direction} for \protect\subref{fig:HC_FM} FM and \protect\subref{fig:HC_AFM} AFM couplings. The magnetization, the flux through the honeycomb plaquettes $W_{\protect\hexagon{0.6}}$ and the NN correlations $\langle S^z_iS^z_j \rangle_z$ are shown.} 
\label{fig:HCsummary}
\end{figure}

For completeness, we provide a brief overview of the results for the honeycomb Kitaev model with both FM and AFM exchange couplings, first reported by \citet{nasu_successive_2018} and summarized in \figref{fig:HCsummary}. 

\paragraph{FM couplings} For FM interactions, switching on the field induces a rapidly increasing magnetization. The Dirac cones are not gapped out but move away from the high symmetry points. At a critical field value $h_c = 0.042$, the magnetization and the flux $W_{\hexagon{0.5}}$ show a discontinuous jump denoting a first-order phase transition from the KSL to the trivial, gapped field-polarized phase. 

\paragraph{AFM couplings} As displayed in \figref{fig:HC_AFM}, the situation is more complex for the AFM Kitaev model.
A continuous transition at $h_{c} \!=\! 0.417$ from the gapless spin liquid to an intermediate phase is followed by a jump in the mean-field parameters at $h_{p} \!=\! 0.503$ when the field-polarized phase is entered. The intermediate phase is also gapless, with both the KSL and the intermediate phase exhibiting a pair of Dirac cones. Precisely at $h_{c}$, a line node connecting the two cones appears, marking the transition. 

As before, the existence of the intermediate phase can be tied to a sign change in an \textit{effective} parameter $E_z^\text{eff}(h)$. 
The exact same shift in $\vect{k}$-space of the Dirac cones and the nodal line that appears when tuning $E_z^\text{eff}(h)$ from negative to positive values also emerges, in the absence of a field, when tuning $J_z$, and hence $E_z^0$, from negative to positive values (keeping $J_x$, $J_y$ AFM). In \refref{nasu_successive_2018}, it was also shown that the change in the topology of the wave function around the Dirac points as you cross $h_c$ is the same as when you cross $J_z=0$ in the zero-field model.

\begin{figure}
	\begin{tikzpicture} 
	\node[inner sep = 0pt]{\includegraphics{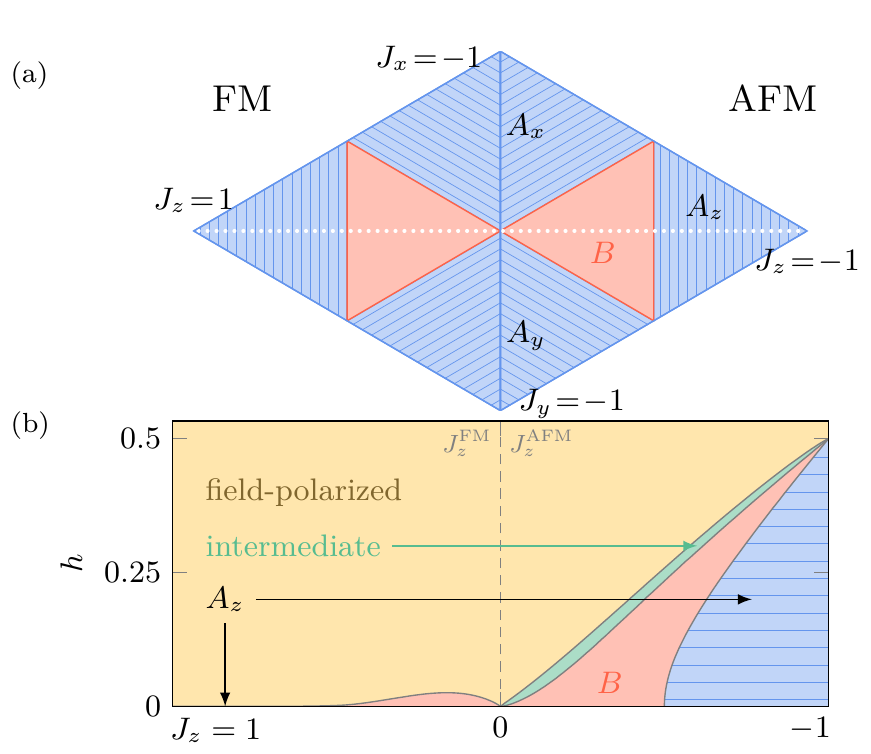}};
	\node[opacity=0]{\subfloat[][\label{fig:HCKitaevPD}]{}};\node[opacity=0]{\subfloat[][\label{fig:HCAnisoPD}]{}};
	\end{tikzpicture} 
	\caption{\textbf{Phase diagram of the Kitaev model} for varying coupling constants with \protect\subref{fig:HCKitaevPD}  $\h=0$ and \protect\subref{fig:HCAnisoPD} for varying field strengths. The coupling constants are taken along the white dotted line sketched in \protect\subref{fig:HCKitaevPD}. $J_x$ and $J_y$ are always AFM, but their signs do not affect the results.} 
\label{fig:HCanisotropies}
\end{figure}

\paragraph{Anisotropic couplings} Going beyond the analysis of \citet{nasu_successive_2018}, we now fix the sum of all absolute values $\sum_i |J_i|\!=\!1$ and consider anisotropies in $J_i$.
To be more specific, $J_z$ is tuned from FM ($J_z\!\!=\!\!1$) to AFM ($J_z\!\!=\!\!-1$), while the couplings $J_x$ and $J_y$ are always fixed to be AFM.
\figref{fig:HCanisotropies} presents the phase diagram as a function of $J_z$ and magnetic field.  The intermediate phase persists for all AFM $J_z$, with $h_c$ naturally increasing as $|J_z|$ increases.
In the limit $J_z \to -1, J_x, J_y \to 0$, the lattice decouples into isolated dimers and the KSL phase is smoothly connected to a covering of dimer singlets on the $z$-bonds with $M^z_i=0$ on all lattice sites, even at finite field, and $E_{\text{singlet}} = -|J_z|/4$ per unit cell. At $\h=1/2$, there is a transition to the polarized phase, with $E = -J_z/4 - \h$.

In the opposite limit, $J_z \to 0$, the width of the intermediate phase shrinks to zero, meaning that the critical field values $h_c,h_p\to 0$. Here, the lattice decouples into one-dimensional chains. The self-consistent solutions for the KSL and polarized phase become identical for $J_z=0, \h=0$ as only the terms $\langle J_{x,y} i c_b c_w \rangle$ contribute equally in both phases to the ground-state energy. A finite $\h$ immediately favors the polarized solution. 

For FM $J_z$, the phase diagram does not host an intermediate phase. The boundary between the KSL and the polarized phase does not monotonically increase as $J_z$ increases, as one might naively expect. As explained above, the KSL and polarized phases are indistinguishable at $J_z\!=\!0$ and $h_p$ shifts to higher values for rising $J_z$ before approaching zero again as $J_z\to +1$. More details on the anisotropic phase diagram are presented in \appref{appendix:MoreResults}.

\begin{figure*}[t]
\begin{tikzpicture}
\node[inner sep =0pt]{\includegraphics{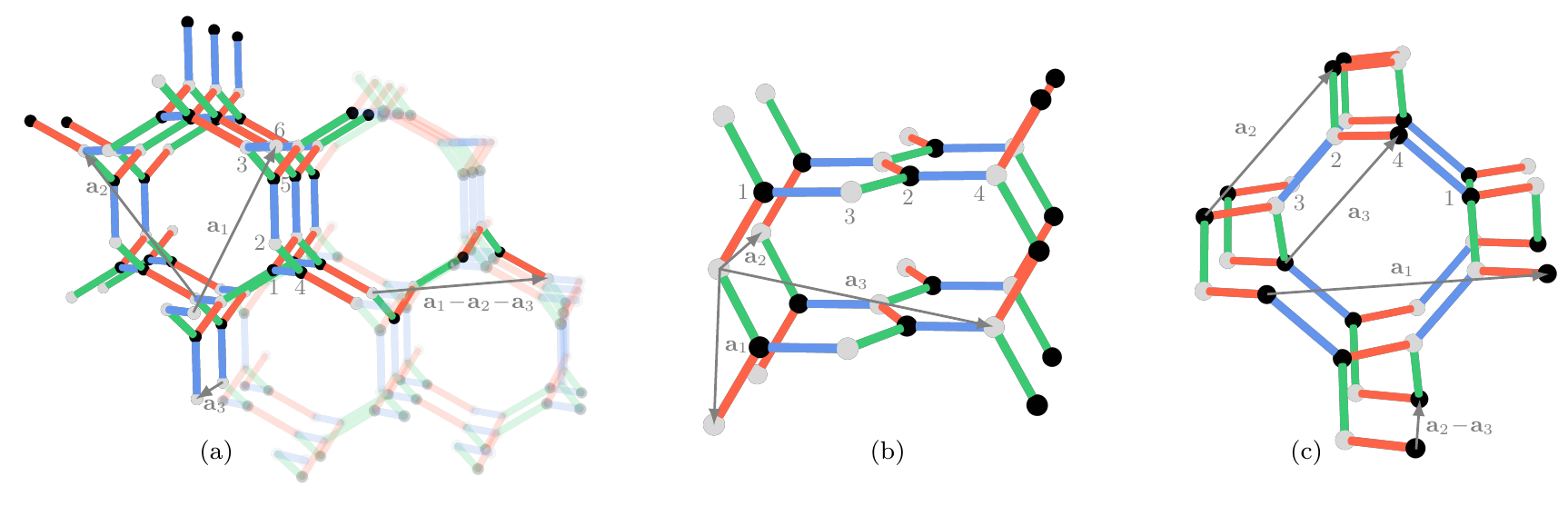}};
\node[opacity=0]{\subfloat[][\label{fig:83b}]{}};
\node[opacity=0]{\subfloat[][\label{fig:103b}]{}};\node[opacity=0]{\subfloat[][\label{fig:103a}]{}};
\end{tikzpicture}
	\caption{\textbf{Unit cells and translation vectors} for the Kitaev model on three dimensional lattices. \protect\subref{fig:83b} For (8,3)b, the Jordan-Wigner chains consist of the $xy$-parts of alternating  up- and down-pointing triangle-spirals, connected by $x$- and $y$-bonds belonging to the dodecagon-spirals and evolve in $\vect{a}_1\!-\!\vect{a}_2\!-\!\vect{a}_3$ direction. \protect\subref{fig:103b} For (10,3)b, the zigzag-chains running in $\vect{a}_1$ and $\vect{a}_2$ direction form the Jordan-Wigner chains. \protect\subref{fig:103a} For (10,3)a, the Jordan-Wigner chains are the $xy$-square spirals and penetrate the lattice in $\vect{a}_2\!-\!\vect{a}_3$ direction.} 
\label{3Dlattices}
\end{figure*}
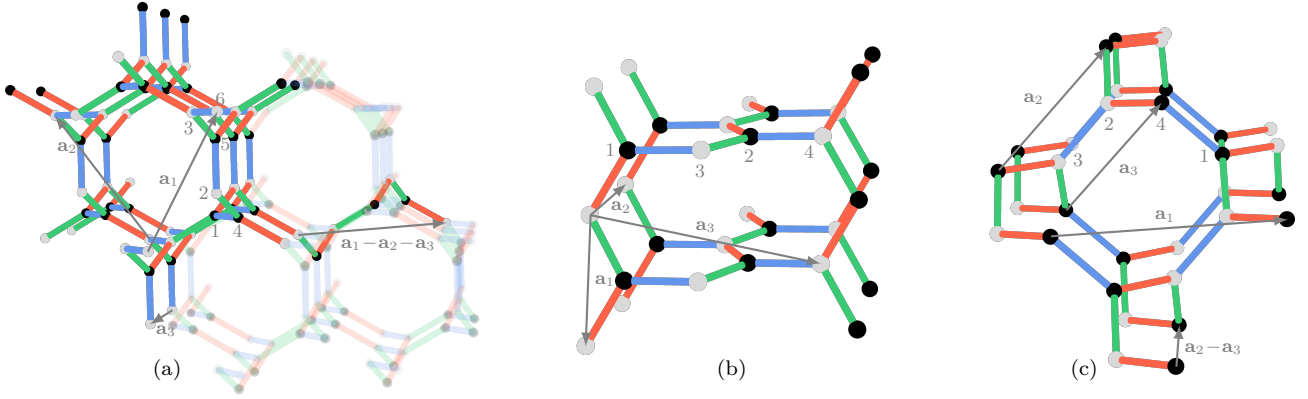

\subsection{\label{sec:3D}3D Kitaev models}
We complete our study of AFM Kitaev spin liquids under the influence of a magnetic field in the [001] direction with a brief discussion of three-dimensional generalization on three representative 3D lattice geometries, shown in \figref{3Dlattices}.
We have selected three different lattice geometries, which are the natural 3D analogues of the 2D lattices of \secref{sec:2D}, that realize Majorana Fermi surfaces with co-dimensions $d_c = 1,2,3$, specifically the lattices denoted as (8,3)b with $d_c=3$, (10,3)b with $d_c=2$ and (10,3a) with $d_c=1$. 

For all of these lattices, the mean-field parameters evolve very similar to the honeycomb lattice case. The codimension has no discernible influence on the stability of the KSL nor on the mean-field parameter evolution. There is a first-order transition to the polarized phase at $h_p$ with an order of magnitude difference between the critical fields for FM and AFM coupling. 
In the latter case, there is also an intermediate phase, which appears at a somewhat smaller field $h_c$. 
The critical fields for all lattices are extremely similar, particularly in the case of AFM couplings, as can be seen in \tabref{table:summary}.
We focus in our discussion on the evolution of the respective Fermi surfaces and point out similarities to their behavior for anisotropic couplings.
The specific form of the mean-field parameters and their exact values at zero field are given in \appref{appendix:MoreResults}.


\subsubsection{(8,3)b lattice}
\begin{figure}[b] 
	\begin{tikzpicture} 
	\node[inner sep = 0pt] {\includegraphics{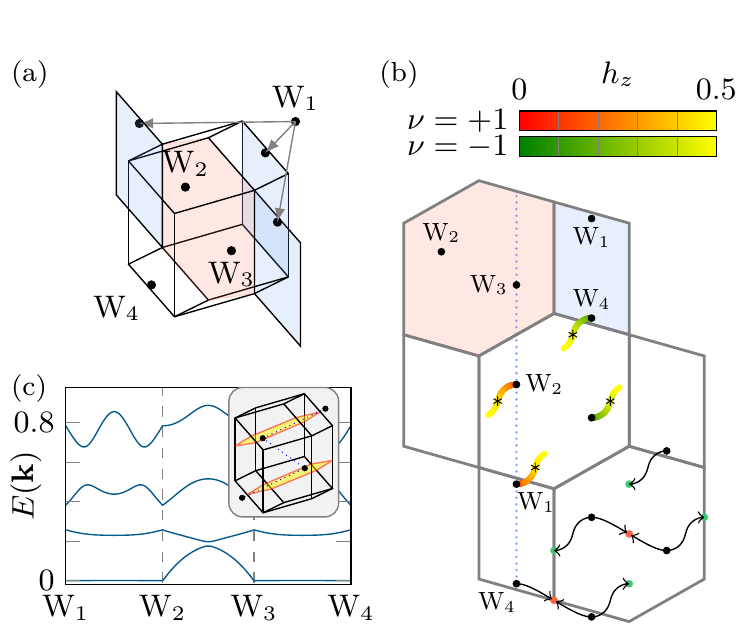}};
	\node[opacity=0]{\subfloat[][\label{fig:83b_1}]{}};\node[opacity=0]{\subfloat[][\label{fig:83b_2}]{}};
	\node[opacity=0]{\subfloat[][\label{fig:83b_3}]{}};
	\end{tikzpicture} 
\caption{\textbf{Brillouin zone, evolution of Weyl points and band structure for (8,3)b.} \protect\subref{fig:83b_1} Brillouin zone and position of Weyl points for isotropic couplings.  \protect\subref{fig:83b_2} Periodic pattern appearing in the extension of the plane marked in red in \protect\subref{fig:83b_1}. The blue dotted line marks the $120^\circ$ invariant line, on which the Weyl points are located for isotropic couplings. The middle hexagon shows the movement of the Weyl points with increasing field. Black stars show the position at $h_{c1}$. For comparison, the evolution for varying FM (towards red dots) and AFM (green dots) $J_z =J_z'$ is shown in the bottom. $J_{x,y}$ is fixed to -1. \protect\subref{fig:83b_3} Energy dispersion at $h_{c2}$. A flat bands occurs in the two planes sketched in yellow in the small inset for $J_z=J_z'=0$. Near the critical fields, the lowest energy band is nearly flat in this plane, as can be seen e.g. between $W_1$ and $W_2$.}
\label{fig:83bsummary}
\end{figure}
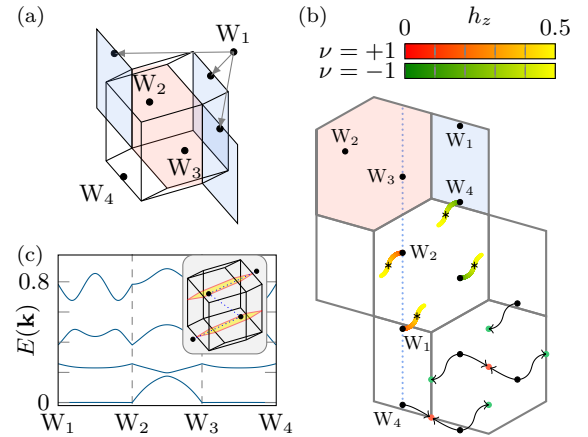

The (8,3)b lattice can probably be best understood as a three-dimensional version of the decorated honeycomb lattice, where the up and down pointing triangles are replaced by coupled counter rotating spirals, see \figref{fig:83b}. There are two different sets of coupling parameters, $J$ for the bonds forming the spirals and $J'$ for the bonds coupling them.
The Kitaev model on this lattice hosts a gapless QSL in a finite parameter region around the isotropic point \cite{obrien_classification_2016}, where the Majorana gap closes with a linear dispersion at isolated points in the Brillouin zone. Such \textit{Weyl points} have been studied in considerable detail in {\em electronic} band structures in the context of Weyl semimetals \cite{wanTopologicalSemimetalFermiarc2011}. As monopoles of Berry curvature, they carry a topological charge. To guarantee overall charge neutrality, Weyl points always occur in pairs of opposite chirality \cite{nielsenAdlerBellJackiwAnomalyWeyl1983}. However, for (8,3)b, a combination of TRS and particle-hole symmetry enforces them to occur in multiples of four \footnote{Note that in contrast to electronic systems, Weyl points occur at the Fermi level even though neither TRS nor inversion symmetry are broken. For a detailed explanation on how the \textit{projective} symmetries make this possible, see \cite{obrien_classification_2016}}. At zero field and for isotropic couplings, we find four Weyl points located in the Brillouin zone as shown in \figref{fig:83b_1}. 

Tuning $J_z$ or $J_z'$ to zero decouples the lattices into effective 2D systems, leading to a vanishing Weyl velocity $v$ and $v'$ in $\vect{k}$-space in the direction normal to these 2D systems. We find that $E_z^{\text{eff}}(h)$ determines the velocity $v$ in precisely the same way as $J_z$ in the zero-field model and similar for $E_{z^\prime}^{\text{eff}}(h)$, $v'$ and $J_z^\prime$. However, $E_z^{\text{eff}}(h)$ and $E_{z^\prime}^{\text{eff}}(h)$ evolve almost identically in field,
and, as a result, the Weyl point evolution is nearly identical to the zero-field case with $J_z = J_z'$. For $J_z\!=\!J_z'=0$, the effective lattice consists of 1D chains and the band structure is therefore dispersionless in planes, as sketched in \figref{fig:83b_3}.  

\paragraph{FM couplings}
As the magnetic field is increased, Weyl points of opposite chirality approach each other, their trajectory mimicking that observed for increasing $J_z=J_z'$ in zero field, sketched at the bottom of \figref{fig:83b_2}. However, while they annihilate in the zero-field anisotropic model at $J_z = 1.51$, for the finite-field isotropic model their movement is terminated by a first-order transition at $h_p=0.030$ to the gapped, polarized phase. In the polarized phase, the magnetization is identical on all lattice sites and all other mean-field parameters vanish.

\paragraph{AFM couplings}
For the AFM case, the evolution of the Weyl points can be compared to the trajectories observed for decreasing $J_z=J_z'$ in zero field, illustrated with the arrows pointing towards the green points in the lower hexagon in \figref{fig:83b_2}. The evolution is almost identical for both cases, driven by the almost identical evolution of $E_z^{\text{eff}}(h)$ and $E_{z^\prime}^{\text{eff}}(h)$. The field-induced movement is color-coded in the middle hexagon. The two velocities $v'$ and $v$ become zero at closely successive values, $h_{c1}\!=\!0.415$ and $h_{c2}\!=\!0.416$, at each of which the spectrum has a line of gapless excitations. At $h_{c1}$, we find $v(h_{c1}) / v(h\!=\!0) \approx 0.0025$ and similar for $v'$ and $h_{c2}$. In the vicinity of the critical fields, the lowest band is thus \textit{nearly} dispersionless in the planes shown in \figref{fig:83b_3}. Details are given in \appref{appendix:MoreResults}. 
Upon further increasing the field, the velocities acquire a finite value again and the Weyl points continue their movement. The polarized phase is entered via a first-order phase transition at $h_p = 0.489$.

\subsubsection{\label{sec:103b}(10,3)b lattice}
\begin{figure} 
 
\begin{tikzpicture} 
	\node[inner sep = 0pt]{\includegraphics{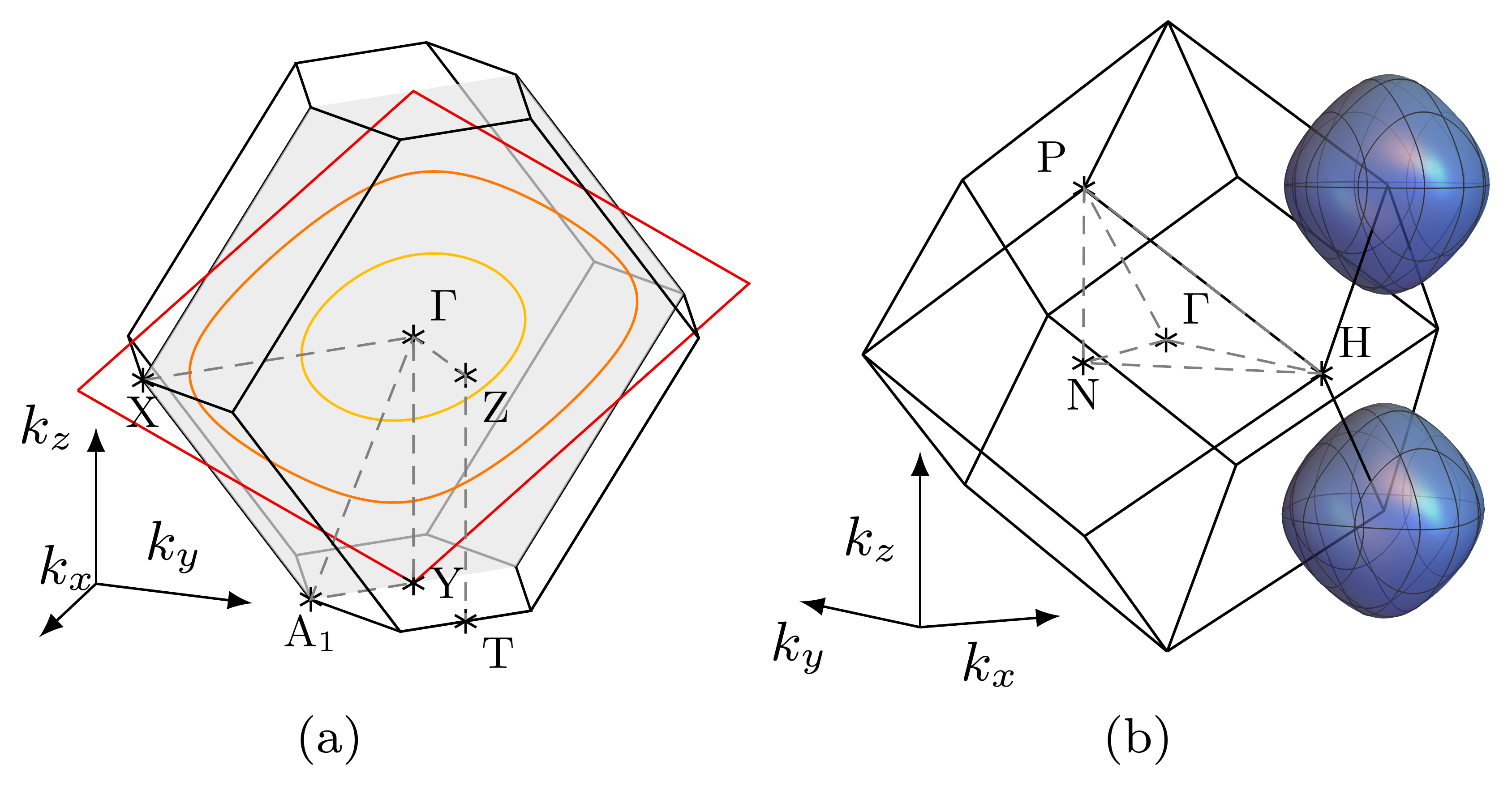}};
	\node[opacity=0.]{\subfloat[][\label{fig:103b-BZ}]{}};\node[opacity=0.]{\subfloat[][\label{fig:103a-BZ}]{}};
\end{tikzpicture} 
\caption{\textbf{Brillouin zone and location of gapless modes} for (10,3)b and (10,3)a. \protect\subref{fig:103b-BZ} For (10,3)b, the gapless modes form a closed line that is shown here for $J_z=1.5$ (yellow), $1$ (orange) and 0 (red). As $J_x=J_y =1$, the gapless line is always located in the plane $k_x=-k_y$, that is grayed out here. \protect\subref{fig:103a-BZ} For (10,3)a, there are two Fermi surfaces located around the corners of the Brillouin zone $(\pi,\pi,\pm\pi)$, shown here for isotropic couplings.}
\label{fig:10_BZ}
\end{figure}

The (10,3)b or hyperhoneycomb lattice is most clearly visualized when it is understood as parallel zigzag chains consisting of $x$- and $y$-bonds along two distinct directions that are coupled by $z$-bonds, see \figref{fig:103b}. This lattice geometry has been widely studied after the discovery that the spin-orbit entangled Mott insulator  $\beta$-Li$_2$IrO$_3$ \cite{takayama_hyperhoneycomb_2015} realizes this structure \cite{kimchi_three-dimensional_2014,lee_heisenberg-kitaev_2014,kimchi_unified_2015,mandal_exactly_2009,hermanns_weyl_2015}. 
The phase diagram of the Kitaev model on (10,3)b looks identical to the honeycomb lattice. Around the point of isotropic couplings, the system hosts a gapless phase with a closed nodal line of zero energy excitations in the Brillouin zone, pictured in \figref{fig:103b-BZ}. This Majorana nodal line lies in the plane $k_x=-k_y$ for $J_x=J_y$ and is protected by a combination of TRS and particle-hole symmetry. For decreasing $J_z$, the nodal line starts shrinking and vanishes at the transition to the gapped spin liquid. As $J_z$ increases, the nodal line expands as shown in \figref{fig:103b-BZ}.
Breaking TRS by applying a field in [111] direction in the perturbative regime gaps out the line everywhere except for two isolated points, giving rise to a Weyl spin liquid phase \cite{hermanns_weyl_2015}.

\paragraph{FM couplings}
For FM couplings and a field in the [001] direction, the nodal line is preserved and starts to shrink. This contraction is terminated by a first-order phase transition at $h_p\! =\! 0.028$ to the gapped, polarized phase. Note that -- in contrast to a [111] field in the perturbative regime -- the nodal line does not gap out, even though TRS is broken by the magnetic field, see \appref{appendix:MoreResults} for details.

\begin{figure}
	\hspace*{-0.5cm}
	\begin{tikzpicture} 
	\node at (0,0){\includegraphics[trim = 0.0cm 0cm 0cm 0cm, clip = true]{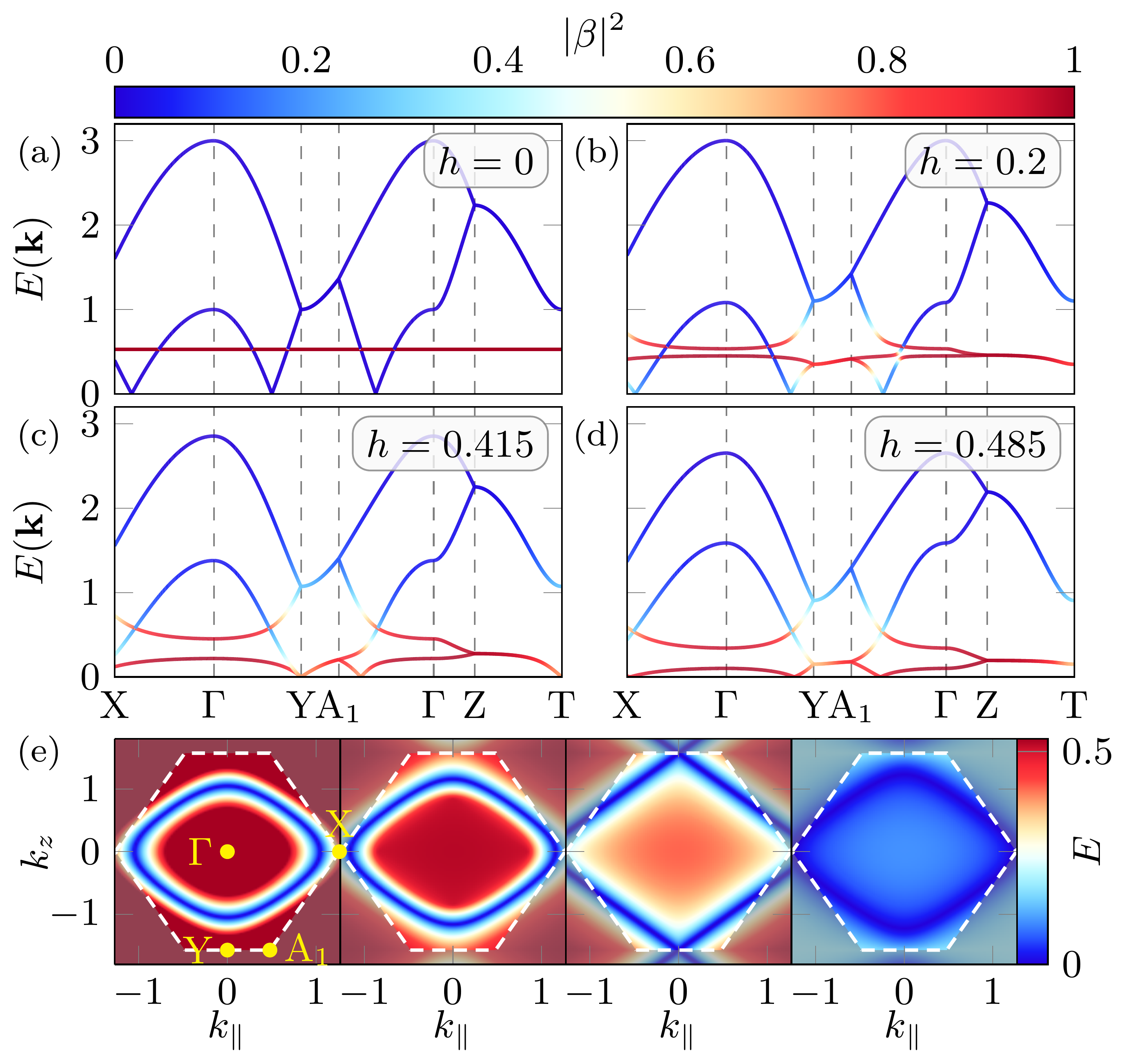}};
	\node[opacity=0.]{\subfloat[][\label{fig:103b-bands1}]{}};\node[opacity=0.]{\subfloat[][\label{fig:103b-bands2}]{}};
	\node[opacity=0.]{\subfloat[][\label{fig:103b-bands3}]{}};\node[opacity=0.]{\subfloat[][\label{fig:103b-bands4}]{}};
	\node[opacity=0.]{\subfloat[][\label{fig:103b-surf}]{}};	
	\end{tikzpicture}
	\caption{\textbf{Mean-field band structure} and color-coded quasiparticle weights for the hyperhoneycomb lattice (10,3)b, with $\beta$ the norm of the $\bar{c}$ part of the wavefunction. \protect \subref{fig:103b-bands1} For $\h\!=\!0$, there are two flat bands, coming from localized $\bar{c}$-Majoranas. \protect \subref{fig:103b-bands2} As $\h$ increases, $c$- and $\bar{c}$-Majoranas mix and the bands hybridize.  \protect \subref{fig:103b-bands3} At $h_c$, there is an additional gap closing coming from the second band. \protect\subref{fig:103b-bands4} Upon further increasing the field, the gapless node shrinks again. \protect \subref{fig:103b-surf} Plot of the lowest lying band in the plane indicated  in gray in \figref{fig:103b-BZ}. The fields are chosen as in the figures above. At $h_c$, the topology of the nodal line changes.}
	\label{fig:103b-bands}
\end{figure}

\paragraph{AFM couplings}
For the AFM case, the nodal line is inflated and at $h_{c}=0.415$, it connects with nodal lines from neighboring Brillouin zones, leading to a change in the topology of the Fermi line, as depicted in \figref{fig:103b-surf}. This transition is accompanied by a closing of the gap for the second band, such that the nodal line at $h_c$ is twofold degenerate, as shown in \figref{fig:103b-bands3}. The spectrum of the zero-energy modes at the transition is again identical to that of the zero-field anisotropic case with $J_z=0$. 
For fields between $h_{c}$ and $h_{p}$, the Fermi line shrinks again. The first-order phase transition to the polarized phase occurs at $h_{p} \!=\! 0.485 $. As the color-coded weight of the $\bar{c}$ contribution to the wave functions  in \figref{fig:103b-bands} shows, $c$- and $\bar{c}$-Majorana sector are separated for $h\!=\!0$ and hybridize for $h\!>\!0$, see also \secref{subsec:SO}.

The similarity between changing $\h$ and $J_z$ can be further substantiated by a simple analytical calculation.
We start with the pure Kitaev model ($\h=0$) and set $J_x = J_y = -1$, but allow for different values of $J_z$. Gapless modes appear for 
\begin{align}
	k_x &= -k_y, \nonumber \\
	\cos(2k_z) &= \frac{J_z^2}{2} - \cos(2 k_x).
	\label{103bgaplessaniso}
\end{align}
On the other hand, the mean-field solution for the isotropic model, with $J_\gamma=J$, under a magnetic field has the structure
\begin{align}
	M_1^z = M_2^z = M_3^z = M_4^z, \nonumber \\
	A_{13} = A_{14} < 0, \nonumber \\
	\bar{A}_{13} = \bar{A}_{14} > 0.
	\label{eq:103bmf}
\end{align}
Defining
\begin{equation}
\xi = \frac{J^2 A \bar{A} + (2J M^z+ 2\h)^2}{J A} ,
\end{equation}
and exploiting the simple form of the mean-field solutions in \equref{eq:103bmf}, the second condition in \equref{103bgaplessaniso} becomes
\begin{equation}
\cos(2k_z) = \frac{\xi^2}{2} - \cos(2k_x)
\label{103bgaplessfield}
\end{equation}
for the gapless modes. If we again define $E_z^\text{eff}(h)$ as in \equref{eq:Ezeff} then $\xi = 4 E_z^\text{eff} / A$ or, more generally, $\xi = J E_z^\text{eff} / E_z^0$. Comparing \equref{103bgaplessaniso} and \equref{103bgaplessfield} directly reveals that the effect of tuning $J_z$ from AFM to FM or changing the sign of $E_z^\text{eff}(h)$ has in fact the \textit{exact} same effect on the nodal lines.
Indeed, $E_z^\text{eff}(h)$ crosses zero at $h_{c}$.


\begin{figure}[t] 
\begin{tikzpicture}
	\node[inner sep = 0pt]{\includegraphics{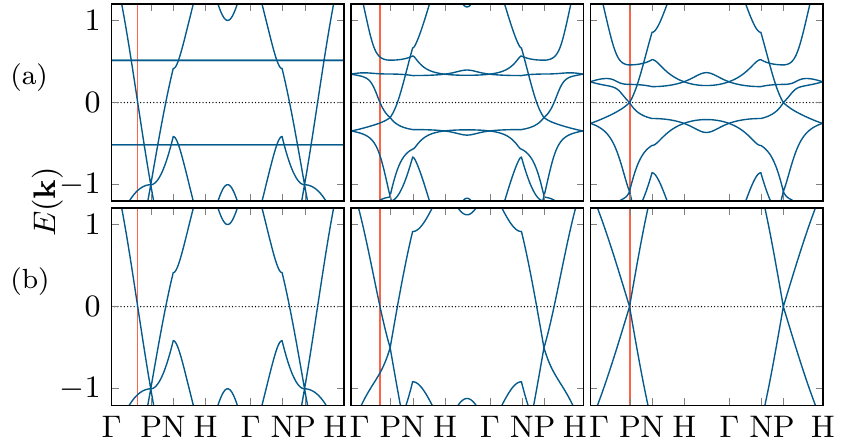}};
	\node[opacity=0.]{\subfloat[][\label{fig:103a-bands-h}]{}};	\node[opacity=0.]{\subfloat[][\label{fig:103a-bands-J}]{}};
\end{tikzpicture} 
\caption{\textbf{Comparison of the (10,3)a band structure} for \protect\subref{fig:103a-bands-h} the isotropic Kitaev model in a magnetic field and \protect\subref{fig:103a-bands-J} the pure Kitaev model with $J_x=J_y$ and varying $J_z$. The parameters here are (from left to right) $\h=0, 0.304,0.416$ and $J_z=-1,-0.5,0$. As the red line indicates, the Fermi surface undergoes the same modifications in both cases, even though the band structures differ. The points along which the spectrum is shown are indicated in \figref{fig:103a-BZ}.}
\label{fig:103a_bands}
\end{figure} 

\subsubsection{(10,3)a lattice}
The (10,3)a or hyperoctagon lattice consists of counter-rotating spirals that are formed by squares and octagons, as illustrated in \figref{fig:103a}. It can be interpreted as a higher-dimensional variant of the square-octagon lattice. It hosts a gapless QSL with two Majorana Fermi surfaces at the point of isotropic couplings \cite{hermanns_quantum_2014}, with \figref{fig:103a-BZ} showing the gapless modes located around the corners of the first Brillouin zone at $(\pi, \pi, \pm \pi)$. As discussed in detail in \refref{obrien_classification_2016}, the Fermi surface for isotropic couplings is topological, with the topological features inherited from the enclosed Weyl points occurring at finite energy. 
\paragraph{FM couplings} 
In the FM Kitaev model, the Fermi surface is stretched in the $k_z$-direction and becomes narrower in the $k_y$- and $k_z$-directions, just as in the case of increasing coupling $J_z$. 
A first-order phase transition to the gapped polarized phase occurs at $h_p \!=\! 0.028$.
\paragraph{AFM couplings} 
For AFM interactions, increasing the field flattens the Fermi surface in the $k_z$-direction. Opposite sides of the Fermi surface approach each other and touch at $h_c \!=\! 0.416$, forming two flat planes. Again, we find that the system undergoes the same Lifshitz transition as when decoupling the lattice by setting $J_z=0$. \figref{fig:103a_bands} illustrates this behavior by comparing the band structure of the isotropic model for different field strengths, \figref{fig:103a-bands-h}, with the zero-field anisotropic case, \figref{fig:103a-bands-J}. Although the band structure differs substantially, the zero-energy modes appear at the same $\vect{k}$-points. In \appref{appendix:MoreResults}, we show analytically that changing the coupling or the magnetic field deforms the Fermi surface in an equivalent way.


\section{\label{sec:Sum}Summary \& outlook}

To summarize, we studied the stability of Kitaev spin liquids in the presence of a uniform magnetic field in the [001] direction for AFM Kitaev models on various two and three dimensional lattice geometries. 
Using a Majorana mean-field approach, based on a Jordan-Wigner representation of the spin degrees of freedom, we mapped out the phase diagrams for both FM and AFM couplings. 
In both cases, a generic phase diagram is obtained, independent of the underlying lattice.
While FM couplings generally result in a single phase transition to the polarized phase at a relatively small field strength $h_p\approx 0.05$, 
the AFM model exhibits a more multifarious phase diagram. 
The KSL is stable up to a critical field strength $h_c$ which is an order of magnitude higher than that of the FM case. At this critical field, the Majorana Fermi surface changes its topology at the transition to an intermediate spin liquid phase.
Interestingly, the gapless modes at the critical field are the same as those that emerge at the zero-field decoupled point, $J_z=0$ and hence $E_z^0=0$. 
Indeed, we showed that the transition can be understood as the vanishing of the effective parameter $E_z^\text{eff}(h)$, defined in \equref{eq:Ezeff}, that occurs when the effect of the external field competes with that of the AFM couplings. The physics due to the sign change of $E_z^\text{eff}(h)$ is directly related to the physics due to the sign change of $E_z^0$ in the zero-field model. 

Our mean-field approach is well suited to investigate the physics of the Kitaev model in the presence of a magnetic field from the view point of the emergent itinerant Majoranas but it has two important limitations. 
The Jordan-Wigner representation does not allow us to study the effects of other field directions. Adding additional magnetic field terms, $-h_{\alpha} \sum_i S_i^{\alpha}$ with $\alpha=x,y$, will lead to highly non-local terms appearing in the fermionized Jordan-Wigner Hamiltonian as $S_i^x$ and $S_i^y$ contain string operators. It is thus not possible to construct a local fermionized Hamiltonian when more than one field component is non-zero, limiting our study to a [001] field. Unlike the conventional parton decompositions used to describe QSLs, the Jordan-Wigner representation does not create an enlarged Hilbert space, nor does it introduce any gauge redundancy. This means that gauge fields, either confined or deconfined, do not naturally appear in our approach. These restrictions imply that our approach cannot capture phase transitions that take place in the gauge sector, as e.g. the Higgs transition between $\mathbb{Z}_2$ and $U(1)$ gauge fields suggested for the honeycomb lattice in a [111] field \cite{hickey_emergence_2019}.

Our study highlights the richness of Kitaev systems with {\em antiferromagnetic} bond-directional exchanges in the presence of a magnetic field, as opposed to their ferromagnetic counterparts. 
The enhanced stability of the KSL and the possibility of realizing transitions between different spin liquid phases, which lie beyond the conventional Landau paradigm, seen in a number of tricoordinated lattice geometries in both two and three spatial dimensions, provides further motivation for the search of new Kitaev materials that naturally exhibit dominant AFM Kitaev interactions. 
Potential routes include exploring Mott materials whose magnetism arises from electrons with an $4f^1$ configuration or whose lattice exhibits polar asymmetry, both of which have been predicted to give rise to AFM Kitaev-type bond-directional exchanges \cite{jang_antiferromagnetic_2019, motome_materials_2020, sugita_antiferromagnetic_2019}. 
An alternative route might be to explore Kitaev materials with higher spin degrees of freedom. Of particular interest might be spin-1 Kitaev systems, which have recently been shown to exhibit much of the same phenomenology as their spin-1/2 counterparts in the presence of a [111] field -- including a strongly enhanced stability for AFM couplings and the occurrence of intermediate spin liquid phases \cite{lee2019tensor,dong2019spin1,zhu2020magnetic,khait2020characterizing,hickey_field-driven_2020}, while experimental realizations are predicted to naturally favor the formation of AFM couplings \cite{Stavropoulos2019}.

	\begin{acknowledgments}
		We acknowledge partial support from the Deutsche Forschungsgemeinschaft (DFG,  German  Research  Foundation),  
		Projektnummer  277146847  --  CRC 1238 (project C03). 
		The numerical simulations were performed on the CHEOPS cluster at RRZK Cologne.
	\end{acknowledgments}
	
	%
	
	\appendix 
	

\section{\label{appendix:Hamiltonian} Explicit expressions for the Hamiltonians}
In this first appendix, we provide explicit expressions for the Kitaev model on all lattices under consideration. Both, the original spin Hamiltonian and its Majorana version after Jordan-Wigner transformation are given. For the three-dimensional lattices, information on unit cells and lattice vectors is also specified.

\subsection{\label{appendix:2D} 2D lattices}

\paragraph{Square-octagon lattice}
The Kitaev model on the square-octagon lattice reads \cite{yang_mosaic_2007}
\begin{align}
H_\text{Kitaev} = &-J_x \sum \limits_{\R} S_{\R-\vect{a}_1,4}^x S_{\R,1}^x  + S_{\R,2}^x  S_{\R-\vect{a}_2,3}^x \nonumber \\  & -J_y \sum \limits_{\R} S_{\R,1}^y S_{\R,2}^y + S_{\R,3}^y  S_{\R,4}^y  \nonumber \\
&	-J_z \sum \limits_{\R} S_{\R,1}^z S_{\R, 3}^z + S_{\R,2}^z S_{\R,4}^z , 
\end{align}
where the sum is taken over all unit cells.
After applying the Jordan-Wigner transformation and adding the magnetic field $h$, we obtain the expression
\begin{align} 
H =&  \frac{i}{4} \sum \limits_{\R} J_x \left[ c_{\R-\vect{a}_1,4} c_{\R,1}  + c_{\R,2}  c_{\R-\vect{a}_2,3} \right] \nonumber \\ & -{J_y} \left[ c_{\R,1} c_{\R,2}  + c_{\R,3} c_{\R,4}  \right] \nonumber \\
& +J_z  \left[i \bar{c}_{\R,1} \bar{c}_{\R,3}  c_{\R,1} c_{\R,3}   + i\bar{c}_{\R,2}  \bar{c}_{\R,4} c_{\R,2} c_{\R,4}   \right]	\nonumber  \\ 
&+2\h \left[ c_{\R,1} \bar{c}_{\R,1} + c_{\R,3} \bar{c}_{\R,3} - c_{\R,2} \bar{c}_{\R,2} -c_{\R,4} \bar{c}_{\R,4}  \right] .
\end{align} 
The black sublattice consists of the lattice sites 2 and 4.

\paragraph{Decorated honeycomb lattice}
The Kitaev model on the decorated honeycomb lattice, introduced by \citet{yao_exact_2007}, is given by
\begin{align} 
H_\text{Kitaev} = &-J_x \sum \limits_{\R} S_{\R,3}^x  S_{\R,2}^x  + S_{\R,5}^x S_{\R,6}^x -J_x' \sum \limits_{\R} S_{\R,4}^x S_{\R+\vect{a}_2,1}^x \nonumber \\
&	-J_y \sum \limits_{\R} S_{\R,1}^y S_{\R,3}^y  +  S_{\R,6}^y  S_{\R,4}^y  - J_y' \sum \limits_{\R} S_{\R,2}^y S_{\R+\vect{a}_1,5}^y \nonumber \\
&	-J_z \sum \limits_{\R} S_{\R,1}^z S_{\R,2}^z  + S_{\R,5}^z  S_{\R,4}^z  -J_z' \sum \limits_{\R} S_{\R,3}^z  S_{\R,6}^z.
\end{align}
Its fermionized version subject to a [001] field is
\begin{align}
H_\text{Kitaev} = & \frac{i}{4} \sum \limits_{\vect{r}} J_x \left[ c_{\vect{r},3}   c_{\vect{r},2}  + c_{\vect{r},5}   c_{\vect{r},6}  \right] +J_x'  c_{\vect{r},4}  c_{\vect{r}+\vect{a}_2,1}    \nonumber \\
& -J_y \left[ c_{\vect{r},1}   c_{\vect{r},3}  +  c_{\vect{r},6}   c_{\vect{r},4}  \right] - J_y' c_{\vect{r},2}  c_{\vect{r}+\vect{a}_1,5} \nonumber \\
&	+J_z \left[i \bar{c}_{\vect{r},1}   \bar{c}_{\vect{r},2}  c_{\vect{r},1}   c_{\vect{r},2}   + i \bar{c}_{\vect{r},4}   \bar{c}_{\vect{r},5}  c_{\vect{r},4}   c_{\vect{r},5}   \right] \nonumber \\
& -J_z' i \bar{c}_{\vect{r},3}   \bar{c}_{\vect{r},6}  c_{\vect{r},3}   c_{\vect{r},6}   \nonumber \\
& + 2\h \left[  c_{\vect{r},1}   \bar{c}_{\vect{r},1}   + c_{\vect{r},2}   \bar{c}_{\vect{r},2}   - c_{\vect{r},3}   \bar{c}_{\vect{r},3}  \right. \nonumber \\ 
&  \left. - c_{\vect{r},4}   \bar{c}_{\vect{r},4}  - c_{\vect{r},5}   \bar{c}_{\vect{r},5}   + c_{\vect{r},6}   \bar{c}_{\vect{r},6}    \right].
\label{eq:Majoranaham_YK}
\end{align}
The black sublattice consists of the lattice sites 3,4,5.

\paragraph{\label{appendix:HC} Honeycomb lattice}
Kitaev's original honeycomb model is given by \cite{kitaev_anyons_2006}
\begin{align}
	H_\text{Kitaev} =& \sum \limits_{\R} -J_x S_{\R-\vect{a}_1,2}^x S_{\R,1}^x  -J_y S_{\R,1}^y S_{\R-\vect{a}_2,2}^y \nonumber \\
	&-J_z S_{\R,1}^z S_{\R,2}^z.
\end{align}
Adding the [001] field and performing the Jordan-Wigner transformation yields \cite{feng_topological_2007,chen_exact_2008,nasu_successive_2018}
\begin{align}	
	H =&  \frac{i}{4} \sum \limits_{\R} J_x c_{\R-\vect{a}_1,2} c_{\R,1} - J_y c_{\R,1} c_{\R-\vect{a}_2,2} \nonumber \\ 
	& - J_z i \bar{c}_{\R,1} \bar{c}_{\R,2} c_{\R,1} c_{\R,2} 
	+2\h \left[ c_{\R,1} \bar{c}_{\R,1} -  c_{\R,2} \bar{c}_{\R,2} \right].
\end{align} 

\subsection{3D lattices}

\paragraph{(8,3)b lattice}
The lattice vectors are chosen as 
\begin{align}
&\vect{a}_1 = \left( \frac{1}{2}, \frac{1}{2\sqrt{3}}, \frac{1}{5} \sqrt{\frac{2}{3}} \right), && \vect{a}_2 = \left( 0, \frac{1}{\sqrt{3}} ,  \frac{2}{5} \sqrt{\frac{2}{3}} \right), \nonumber \\
& \vect{a}_3 = \left( 0,0,\frac{\sqrt{6}}{5} \right).
\label{83bvecs}
\end{align}
The lattice (8,3)b has six sites per unit cell at the positions
\begin{align}
&	\vect{r}_1 = \left( \frac{1}{10}, \frac{1}{2\sqrt{3}}, \frac{1}{5} \sqrt{\frac{2}{3}} \right), && \vect{r}_2 = \left( \frac{1}{5}, \frac{\sqrt{3}}{5}, \frac{\sqrt{6}}{5} \right), \nonumber \\
& \vect{r}_3 =  \left( \frac{3}{10} ,\frac{11}{10\sqrt{3}}, \frac{4}{5} \sqrt{\frac{2}{3}}\right),  &&	\vect{r}_4 = \left( \frac{1}{5}, \frac{2}{5\sqrt{3}}, \frac{2}{5} \sqrt{\frac{2}{3}} \right), \nonumber \\
& \vect{r}_5 = \left( \frac{3}{10}, \frac{3\sqrt{3}}{10}, \frac{\sqrt{6}}{5} \right),  && \vect{r}_6 =  \left( \frac{2}{5}, \frac{1}{\sqrt{3}} , \sqrt{\frac{2}{3}}\right).
\end{align}
The Kitaev model then reads 
\begin{align}
H_\text{Kitaev} = &-J_x \sum \limits_{\R} S_{\R,5}^x  S_{\R-\vect{a}_3,6}^x + S_{\R,1}^x  S_{\R-\vect{a}_3 ,2}^x \nonumber \\ 
& -J_x' \sum \limits_{\R} S_{\R,4}^x  S_{\R -\vect{a}_2,3}^x 	-J_y \sum \limits_{\R} S_{\R,3}^y  S_{\R,5}^y  +  S_{\R,2}^y  S_{\R,4}^y \nonumber \\ 
& - J_y' \sum \limits_{\R} S_{\R-\vect{a}_3-\vect{a}_1,6}^y  S_{\R,1}^y  \nonumber \\
&	-J_z \sum \limits_{\R} S_{\R,1}^z  S_{\R,4}^z   + S_{\R,3}^z  S_{\R,6}^z -J_z' \sum \limits_{\R} S_{\R,2}^z  S_{\R,5}^z .
\end{align}
The Jordan-Wigner chains run along the direction $\vect{a}_1\!-\!\vect{a}_2\!-\!\vect{a}_3 $. They are visualized in \figref{fig:83b}.
Expressed in terms of Jordan-Wigner fermions and after adding the magnetic field, the Hamiltonian becomes
\begin{align} 
H=& \frac{i}{4} \sum \limits_{\R} J_x \left[  c_{\R,5} c_{\R-\vect{a}_3,6} + c_{\R,1}  c_{\R -\vect{a}_3,2} \right]  + J'_x c_{\R,4} c_{\R-\vect{a}_2,3} \nonumber \\ 
& -J_y \left[ c_{\R,3}  c_{\R,5}  + c_{\R,2} c_{\R,4}  \right] + J'_y c_{\R-\vect{a}_3-\vect{a}_1,6} c_{\R,1} \nonumber \\
& +J_z \left[i \bar{c}_{\R,1} \bar{c}_{\R,4}  c_{\R,1} c_{\R,4}  + i \bar{c}_{\R,3} \bar{c}_{\R,6}  c_{\R,3} c_{\R,6} \right] \nonumber \\
& - J'_z i \bar{c}_{\R,2} \bar{c}_{\R,5}  c_{\R,2} c_{\R,5}  \nonumber \\
& +2\h \left[ c_{\R,2}\bar{c}_{\R,2} + c_{\R,3} \bar{c}_{\R,3} + c_{\R,6} \bar{c}_{\R,6} \right. \nonumber \\ 
& \left. - c_{\R,1} \bar{c}_{\R,1} - c_{\R,4} \bar{c}_{\R,4} - c_{\R,5} \bar{c}_{\R,5} \right].  
\end{align} 
The black sublattice consists of the lattice sites 1, 4 and 5.

\paragraph{(10,3)b lattice}
With the lattice vectors given by 
\begin{align}
& \vect{a}_1 = (-1,1,-2), && 	\vect{a}_2 = (-1,1,2), \nonumber \\  &\vect{a}_3 = (2,4,0),
\end{align}
and the four sites per unit cell at
\begin{align}
\vect{r}_1 = (0,0,0), && \vect{r}_2 = (1,2,1), \nonumber \\
\vect{r}_3 = (1,1,0), && \vect{r}_4 = (2,3,1),  
\end{align}
the Kitaev model on the lattice (10,3)b reads
\begin{align}
H_\text{Kitaev} = &-J_x \sum \limits_{\R} S_{\R,1}^x  S_{\R+\vect{a}_1-\vect{a}_3,4}^x  + S_{\R-\vect{a}_2,2}^x  S_{\R,3}^x  \nonumber \\ 
& -J_y \sum \limits_{\R} S_{\R-\vect{a}_3,4}^y  S_{\R,1}^y  + S_{\R,3}^y  S_{\R,2}^y   \nonumber \\
&	-J_z \sum \limits_{\R} S_{\R,1}^z  S_{\R,3}^z  + S_{\R,2}^z  S_{\R,4}^z .
\end{align}
The Jordan-Wigner chains correspond to the two different $xy$-zigzag chains running in positive $\vect{a}_1$($\vect{a}_2$) direction. After adding the magnetic field and transforming to Jordan-Wigner fermions, the Hamiltonian reads
\begin{align}
H =&  \frac{i}{4} \sum \limits_{\R} J_x \left[ c_{\R,1}   c_{\R+\vect{a}_1 - \vect{a}_3,4} +  c_{\R- \vect{a}_2,2}  c_{\R,3} \right] \nonumber \\ 
& - J_y \left[ c_{\R-\vect{a}_3,4} c_{\R,1}  + c_{\R,3} c_{\R,2} \right]  \nonumber \\
& - J_z \left[ i \bar{c}_{\R,1}  \bar{c}_{\R,3} c_{\R,1}  c_{\R,3} + i \bar{c}_{\R,2} \bar{c}_{\R,4} c_{\R,2} c_{\R,4} \right] \nonumber \\ 
& +2\h \left[ -c_{\R,1}  \bar{c}_{\R,1}  -  c_{\R,2} \bar{c}_{\R,2}  +c_{\R,3} \bar{c}_{\R,3} +  c_{\R,4} \bar{c}_{\R,4} \right].
\end{align}
The black sublattice consists of the lattice sites 1 and 2. 

\paragraph{(10,3)a lattice}
The lattice vectors for (10,3)a are given by
\begin{align}
& \vect{a}_1 = (1,0,0), && 	\vect{a}_2 = \frac{1}{2}(1,1,-1), \nonumber \\ & \vect{a}_3 = \frac{1}{2}(1,1,1). 
\end{align}
The unit cell vectors 
\begin{align}
&	\vect{r}_1 = \frac{1}{8}(3,1,1), && \vect{r}_2 = \frac{1}{8}(-1,3,-1),  \nonumber \\
&	\vect{r}_3 = \frac{1}{8}(-3,1,-1), && \vect{r}_4 = \frac{1}{8}(1,3,1),
\end{align}
are chosen such that lattice sites connected by $z$-bonds belong to the same unit cell. The Kitaev model is \cite{hermanns_quantum_2014}
\begin{align}
H_\text{Kitaev} = &-J_x \sum \limits_{\R} S_{\R,4}^x  S_{\R,2}^x  + S_{\R- \vect{a}_1,1}^x  S_{\R,3}^x \nonumber \\
& -J_y \sum \limits_{\R} S_{\R -\vect{a}_2,2}^y  S_{\R -\vect{a}_1,1}^y  + S_{\R,3}^y  S_{\R-\vect{a}_3,4}^y   \nonumber \\
&	-J_z \sum \limits_{\R} S_{\R,1}^z  S_{\R,4}^z + S_{\R,2}^z  S_{\R,3}^z .
\end{align}
The Jordan-Wigner chains correspond to the square spirals running in $\vect{a}_2 \! - \! \vect{a}_3$ direction. The Hamiltonian in the presence of a [001] field is then calculated to 
\begin{align}
H =&  \frac{i}{4} \sum \limits_{\R} J_x \left[  c_{\R,4} c_{\R,2} + c_{\R-\vect{a}_1 ,1}  c_{\R,3} \right] \nonumber \\ 
& - J_y \left[ c_{\R-\vect{a}_2,2} c_{\R-\vect{a}_1 ,1}  + c_{\R,3} c_{\R-\vect{a}_3,4} \right] \nonumber \\
& + J_z \left[ i \bar{c}_{\R,1} \bar{c}_{\R,4} c_{\R,1} c_{\R,4} + i \bar{c}_{\R,2} \bar{c}_{\R,3} c_{\R,2} c_{\R,3} \right] \nonumber \\ 
& +2\h \left[ -c_{\R,1} \bar{c}_{\R,1} +  c_{\R,2} \bar{c}_{\R,2}  +c_{\R,3} \bar{c}_{\R,3} -  c_{\R,4} \bar{c}_{\R,4} \right].
\end{align}
The black sublattice consists of the lattice sites 1 and 4.

\begin{figure*}
	\begin{tikzpicture} 
	\node[inner sep=0pt]{\includegraphics{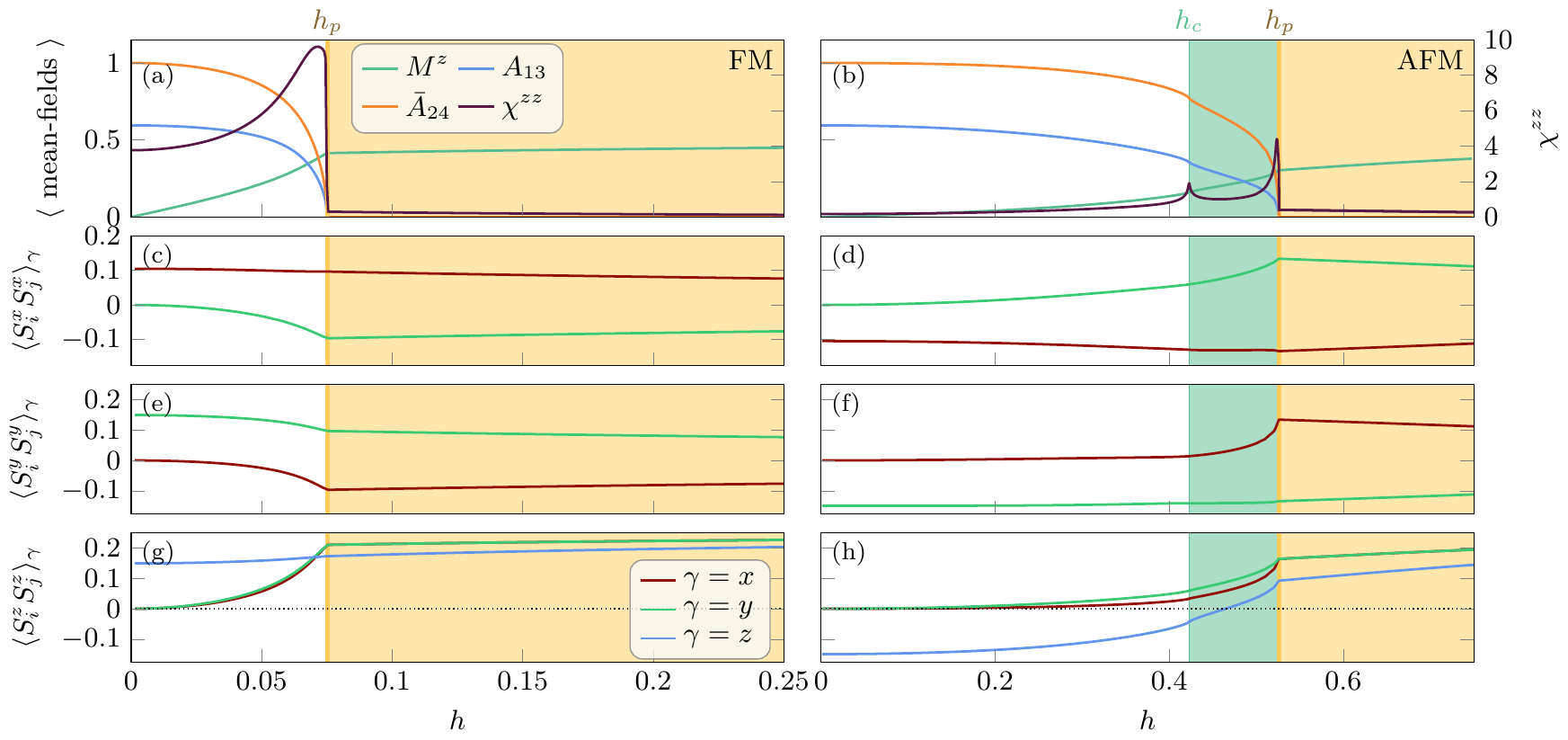}};
	\node[opacity=0.]{\subfloat[][\label{fig:SO_MF_FM}]{}};
	\node[opacity=0.]{\subfloat[][\label{fig:SO_MF_AFM}]{}};
	\node[opacity=0.]{\subfloat[][\label{fig:SO_SxSx_FM}]{}};
	\node[opacity=0.]{\subfloat[][\label{fig:SO_SxSx_AFM}]{}};
	\node[opacity=0.]{\subfloat[][\label{fig:SO_SySy_FM}]{}};;
	\node[opacity=0.]{\subfloat[][\label{fig:SO_SySy_AFM}]{}};;
	\node[opacity=0.]{\subfloat[][\label{fig:SO_SzSz_FM}]{}};;
	\node[opacity=0.]{\subfloat[][\label{fig:SO_SzSz_AFM}]{}};;
	
	\end{tikzpicture}
	\caption{\textbf{Mean-field parameters, susceptibility and correlations for the square-octagon lattice.} Mean-fields and magnetic susceptibility $\chi^{zz}$ for \protect\subref{fig:SO_MF_FM} FM couplings and \protect\subref{fig:SO_MF_AFM} AFM couplings.  
		The field dependence of the correlators $\langle S_i^{\gamma'} S_j^{\gamma'} \rangle$ along the bond $\gamma$ is shown in \protect\subref{fig:SO_SxSx_FM}, \protect\subref{fig:SO_SySy_FM}, \protect\subref{fig:SO_SzSz_FM} for FM couplings and in \protect\subref{fig:SO_SxSx_AFM}, \protect\subref{fig:SO_SySy_AFM}, \protect\subref{fig:SO_SzSz_AFM} for AFM couplings. For $\gamma' \!=\!x,y$ and $\gamma=z$, string factors from the Jordan-Wigner chains emerge and prevent a simple evaluation of the correlation functions.}
	\label{fig_SO_appendix1} 
\end{figure*} 

\section{\label{appendix:MoreResults} Details on the mean-field results}
In this second appendix, we extend the discussion of the mean-field results. For all lattices, we provide the mean-field parameters for $h\!=\!0$, if not already done in the main text. Correlation functions and susceptibility are discussed using the example of the square-octagon lattice. For honeycomb and decorated-honeycomb lattice, more details on the anisotropic phase diagrams are provided. For (8,3)b, the band structure is subjected to closer examination. For (10,3)b, we discuss the stability of the nodal line and for (10,3)a, we calculate and compare the zero-energy modes for the anisotropic model in zero-field and the isotropic model subject to a field.

\subsection{Square-octagon lattice}
\figref{fig:SO_MF_FM} and \figref{fig:SO_MF_AFM} show the evolution of the mean-field parameters and the susceptibility.
The mean-fields evolve continuously as $h_p$ is crossed.
All other lattices (except for decorated honeycomb) show a similar evolution, with the only difference that there the transition at $h_p$ is of first order.
The susceptibility $\chi$ jumps at $h_p$ for both, AFM and FM couplings. For AFM couplings, the kink in the mean-field evolution at $h_c$ manifests itself in a further peak in $\chi$. 
The evolution of the NN spin-spin correlations $\langle S_i^\gamma S_j^\gamma \rangle$ is shown in \figref{fig:SO_SxSx_FM} - \figref{fig:SO_SzSz_AFM}. 
$\langle S_i^{\gamma} S_j^{\gamma} \rangle_{\gamma '}$ contains only terms of the form $\langle c_i c_j \rangle$ ($\langle \bar{c}_i \bar{c}_j \rangle$) for $\gamma \!=\! \gamma'$ ($\gamma \! \neq  \! \gamma'$) and therefore measures the kinetic energy of the $c$- ($\bar{c}$-)Majorana fermions \cite{nasu_successive_2018}. 
The $\bar{c}$-Majoranas, localized on the $z$-bonds for $\h\!=\!0$, start hopping as $\h$ is increased and the corresponding correlations become non-zero. They evolve non-monotonic in $\h$ for $x$- and $y$-bonds with a maximum in the absolute value near $h_p$, see \figref{fig:SO_SxSx_FM} - \figref{fig:SO_SySy_AFM}. Similar results are obtained for the honeycomb lattice \cite{nasu_successive_2018}. 

\subsection{Decorated honeycomb lattice}

\begin{figure} 
	\includegraphics{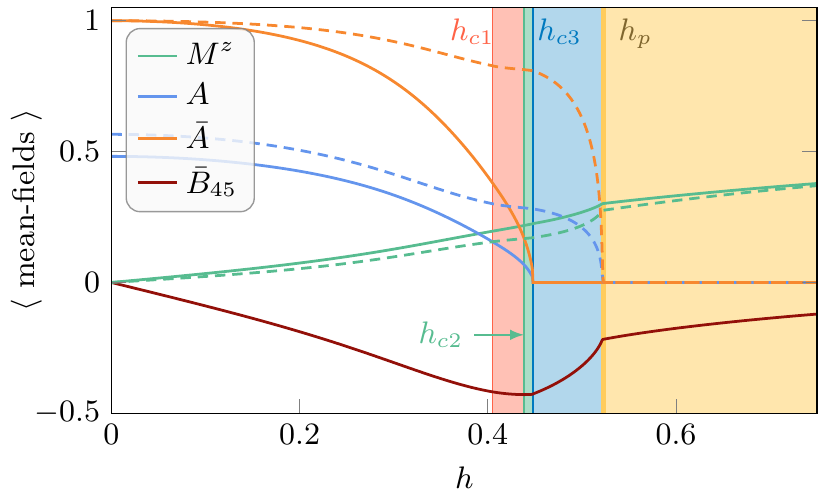} 
		\caption{\textbf{Evolution of the mean-field parameters in the isotropic AFM Kitaev model on the decorated honeycomb lattice}. Solid (dashed) lines denote mean-field parameters belonging to solid (dashed) bonds in \figref{fig:YK}.}
	\label{fig:YKappendix}
\end{figure}

\paragraph{Mean-field parameters for $J\!=\!J'$}
The relative signs of the AFM $A$ and $\bar{A}$ given in \equref{YK_mf_1} change for FM couplings. This holds for all lattices.
\figref{fig:YKappendix} illustrates the evolution of the mean-fields in an increasing field $h$.
Throughout the whole range of $h$, the relations
\begin{align} 
	&M^z \coloneqq  M_1^z\!=\!M_2^z\!=\!M_4^z\!=\!M_5^z \neq M_3^z\!=\!M_6^z \eqqcolon M'^z, \nonumber \\
	&\bar{B}_{12} = -\bar{B}_{45} = B_{12} = -B_{45} \label{YK_mf_appendix}
\end{align}
hold. In addition, for $h\!<\!h_{c3}$, the mean-fields
\begin{align} 
	&   A_{36}, \bar{A}_{36} \neq 0, \nonumber \\
	&	A_{12} = A_{45}, \phantom{=} \bar{A}_{12} = \bar{A}_{45}, 
	\label{YK_mf_hc1}
\end{align}
are non-zero.
For solid (dashed) bonds, $A$ and $\bar{A}$ approach zero at $h_{c3}$ ($h_{c4}$). Between $h_{c3}$ and $h_{c4}$, the phase is thus characterized by \equref{YK_mf_appendix} and
\begin{equation} 
	A_{36}, \bar{A}_{36} \neq 0, 
	\label{YK_mf_hc3}
\end{equation}
In the polarized phase, the mean-fields are given by \equref{YK_mf_appendix} alone.
 $B$ and $\bar{B}$ vanish for $\h \!\to\! \infty$. 
The signatures of the phase transitions  (the `kinks') at $h_{c1}$, $h_{c2}$, albeit difficult to recognize, still occur in \figref{fig:YKappendix}.

\paragraph{Dirac points in the anisotropic Kitaev model}
In order to compare the Dirac points for $J_z\!=\!0$ and $E_z^{\text{eff}}=0$, we calculate their positions for fixed AFM couplings on $x/x'$- and $y/y'$-bonds and $J_z\!=\!0$. 
They are then located at  
\begin{align}
	k_x &= \pm 4 \arctan \left( \frac{-2 + \sign(J_z') J_z' \sqrt{ \frac{4 - J_z'^2}{J_z'^2} }} {J_z'}  \right), \nonumber \\ 
	k_y &= \pm \frac{\pi}{\sqrt{3}}.
\end{align}
For $J_z'=-1$, their position is $\pm (\pi/3, \pi/\sqrt{3})$. As $|J_z'|$ decreases, the gapless points move towards $k_x = 0$, crossing zero as $J_z'$ changes from AFM to FM. This is the same behavior found along the black dashed line in \figref{fig:YKaniso}, where $E_z^{\text{eff}}\!=\!0$. Starting at high $J'$, we find the same gapless excitation that a system with $J_z=0$ and a small $J_z'$ that changes from AFM to FM as the red line -- where $E_{z'}^{\text{eff}}\!=\!0$ -- is crossed, hosts.

\paragraph{Details on the anisotropic phase diagram} Here, we provide more details on the phases appearing in \figref{fig:YKaniso}.

The CSLs (green, red) all share the parameter configuration characterized by \equref{YK_mf_appendix} and \equref{YK_mf_hc1}.
The intermediate phase shown in yellow, which occurs at low $J'$ is called $\vartriangle_1$.
It has the same parameter configuration as the polarized phase, i.e. \equref{YK_mf_appendix}.
The narrow, orange phase, separating the polarized phase and $\vartriangle_1$, is called $\vartriangle_2$. Its mean-field parameters are discussed below.
The intermediate phase highlighted in blue is called $I_1$ and characterized by \equref{YK_mf_appendix} and \equref{YK_mf_hc3}. It exhibits two peculiarities, an intraphase transition marked by the blue dotted line in \figref{fig:YKaniso}, where the mean-fields evolve discontinuously and a kink occurring in the boundary to the CSL at  $J'/J=0.94$, originating in an additional level crossing of the two coexisting self-consistent solutions belonging to $I_1$ and CSL respectively. 

In order to understand the origin of the intermediate phases, we will now address the two limiting cases $J\!=\!0$ and $J'\!=\!0$.

\subparagraph{The limit  \boldmath{$J \! \to \! 0$}}
For $J=0$, the system decouples to isolated dimers. The ground state energy for $x'$- and $y'$-dimers is $E_0\!=\!-\frac{1}{4}\sqrt{1 + 16 \h^2}$ and the magnetization is given by $M^z\! =\! 2\h / \sqrt{1 + 16\h^2}$. 
The mean-field solution $I_1$ reproduces these exact results.
For $z$-bonds, AFM order and a covering of dimer singlets are degenerated, but an arbitrary small $J$ prefers the solution stemming from the latter. For $J\!=\!0$, the transition at $\h_p\!=\!0.5$ only affects the $z'$-bonds, where $M'^z$ changes from zero to fully polarized. 
For $J,h \!=\! 0$, the phases $I_1$ and CSL both have the mean-field parameters $A_{36}\! = \! -\bar{A}_{36} = 1$ and in the CSL phase $\bar{A}_{12} = \bar{A}_{45} = 1$ additionally applies. CSL and $I_1$ are then degenerated because the energy contributions from the solid bonds vanish for $J\!=\!0$. However, the CSL still has $|W|=1$ and AFM correlations on solid bonds for any finite $J$, in contrast to the phase $I_1$ with $|W|=0$ and FM correlations on solid bonds.
To summarize, we find that the intermediate phase $I_1$ is smoothly connected to the exact solution at $J\!=\!0$ and $h\!<\!0.5$, consisting of singlets on $z'$-bonds and singlet formation competing with magnetic order on $x'$- and $y'$-bonds.

\subparagraph{The limit \boldmath{$J' \! \! \to \! 0$}}
For $J'\!=\!0$, the system decouples to a collection of independent triangles. Two eigenenergies of the triangle model are of relevance, $E_1 \!=\! -(\sqrt{3} + 2\h)/4$ and $E_2 \!=\! - (2\h +\Upsilon ) / 4$, where the shorthand notation $\Upsilon = \sqrt{3-8\h+16\h^2}$ was introduced. They are degenerated at $\h\!=\!0$ and intersect again at $\h\!=\!0.5$. In between, $E_1$ is the ground-state energy.
In this field range, the field-independent mean-fields
\begin{align}
M^z = \frac{3 - \sqrt{3}}{12},&& M'^z =  \frac{1}{2 \sqrt{3}}, &&B_{12} = \frac{ 3 \!+\! \sqrt{3}}{6},
\label{YKmagpol1}
\end{align}
are exact and reproduce the correct $E_1$. The correlations $\langle S_i^z S_j^z \rangle_z$ are AFM (FM) for solid (dashed) bonds. The phase $\vartriangle_1$ can be understood as a descendant from the exact solution for $h\!<\!0.5$.
At $\h\! = \!0.5$, the mean-field parameters on solid bonds change discontinuously to  
\begin{align} 
	M^z =  \frac{3 + \sqrt{3}}{12} , &&	B_{12} = \frac{3 - \sqrt{3}}{6},	
	\label{YKmagpol2}
\end{align} 
which induces a sign change in the correlation from AFM to FM.
The magnetizations now evolve according to 
\begin{align}
M^z = \frac{1}{4} + \frac{4\h-1}{4\Upsilon}, && M'^z = \frac{4\h-1}{2 \Upsilon}.
\label{YKmagpol3}
\end{align}
Again, the mean-field solution is exact and reproduces the correct ground-state energy $E_2$. The polarized phase is thus smoothly connected to the exact solution of the triangle system with energy $E_2$.
At $\h =0.5$, the phase $\vartriangle_2$ emerges between $\vartriangle_1$ and the polarized phase.
Its mean-field configuration is similar to the polarized phase, but in addition, the parameters $\bar{A},A$ on solid bonds become non-zero for $J'>0$. For $J\! \! = \!  \!0$, the width of this phase is zero and we find $M^z\!= \! 1/4$, $B_{12}\!=\!1/2$ and $M'^z \! = \! 1/(2\sqrt{3})$. This is again an exact solution, corresponding to a superposition of the states described by \equref{YKmagpol1} and \equref{YKmagpol2}, that are degenerated for $h\!=\!0.5$.

To summarize, we find that the intermediate phases $\vartriangle_1$ and $\vartriangle_2$ at finite $J'$ can both be understood as stemming from exact solutions of the triangle system. The same holds for the polarized phase that approaches an exact solution of the isolated triangles for $J'\to 0$.

\subsection{Honeycomb lattice}
\paragraph{Mean-field parameters}
At zero field and for FM couplings, the mean-field parameters are given by
\begin{align}
	A_{12} = 0.525, && \bar{A}_{12} = 1.
\end{align}
The magnetization on the two sublattices is equal for all $\h$. $B$ and $\bar{B}$ are always zero, as for all lattices except for the decorated honeycomb lattice.
\begin{figure} 
	\begin{tikzpicture}
	\node[inner sep = 0pt]{\includegraphics{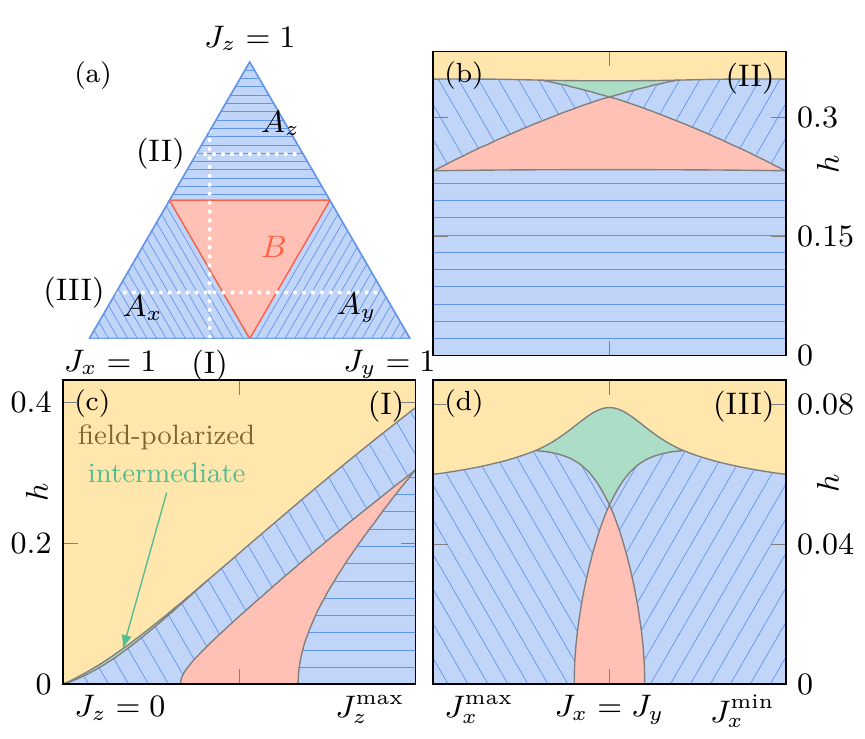}};
	\node[opacity = 0]{\subfloat[][\label{fig:PDKitaev2a}]{}};\node[opacity = 0]{\subfloat[][\label{fig:anisoPathII}]{}};
	\node[opacity = 0]{\subfloat[][\label{fig:anisoPathI}]{}};\node[opacity = 0]{\subfloat[][\label{fig:anisoPathIII}]{}};
	\end{tikzpicture}	
	\caption{\textbf{Phase diagram for anisotropic couplings.} \protect\subref{fig:PDKitaev2a} Phase diagram showing the three pathes along which the anisotropic phase diagrams are plotted in \protect\subref{fig:anisoPathII}-\protect\subref{fig:anisoPathIII}. \protect\subref{fig:anisoPathII} Phase diagram for $J_z=2/3$.  \protect\subref{fig:anisoPathI} Phase diagram that includes the point $J_x\!=\!5/8$, $J_y=3/8$, $J_z=0$.  \protect\subref{fig:anisoPathIII} Phase diagram for $J_z=1/6$. }
	\label{fig:appHCPD}
\end{figure}
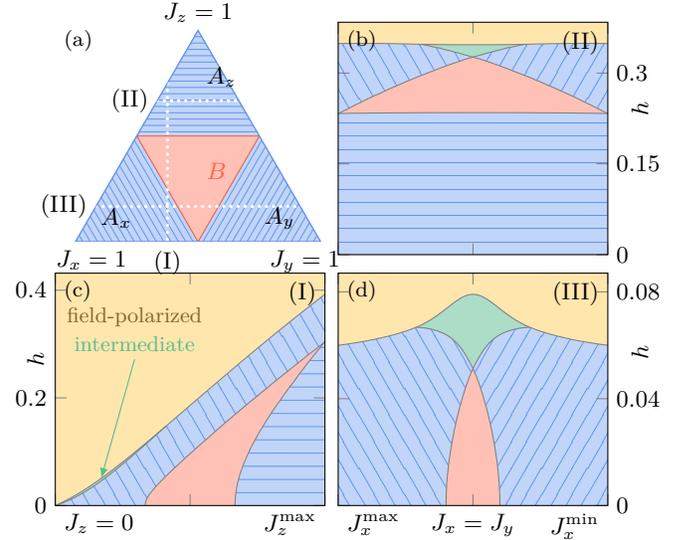

\paragraph{Details on the anisotropic phase diagram}
Here, we discuss the anisotropic Kitaev model along further lines in coupling space, shown in  \figref{fig:PDKitaev2a}.
As can be seen in \figref{fig:anisoPathII} and \figref{fig:anisoPathIII}, intermediate phase and $B$ phase only touch for $J_x=J_y$. For small $J_z$ and large anisotropies between $J_x$ and $J_y$, the gapless intermediate phase appears even if the ground state at $\h\!=\!0$ is the gapped $A_{x/y}$ phase, as shown in \figref{fig:anisoPathIII} and \figref{fig:anisoPathI}.
The nodal line at $h_c$ disappears if mirror symmetries are broken by $J_x\! \neq \!J_y$ \cite{liang_intermediate_2018}.
Instead, the transitions from the $B$ phase to the $A_{x/y}$ phase and then on to the intermediate phase are characterized by an opening and reclosing of the Dirac points.
A further gapped intermediate phase for small anisotropies in $J_{x/y}$, identified here as stemming from the phase $A_{x/y}$, was also found in \refref{liang_intermediate_2018}.

\subsection{(8,3)b}
\paragraph{Mean-field parameters}
For $\h=0$ and FM couplings, the mean-field parameters are given by
\begin{align}
	A_{14} &= -A_{36} = -0.521, & A_{25} &= -0.505, \nonumber \\
	\bar{A}_{14} &= - \bar{A}_{36} = 1,	&	\bar{A}_{25} &= -1.
\end{align}
The relations between the mean-fields also hold for $h\!>\!0$. In the KSL phase, the magnetization $M_2^z,M_5^z$ differ from the rest. In the polarized phase, all $M_i^z$ are equal.

\paragraph{Evolution of Weyl points} 
With \equref{83bvecs} given, the reciprocal lattice vectors are
\begin{align}
	\vect{b}_1 &= \left(4 \pi, 0,0\right), 	& \vect{b}_2 = \left(-2 \pi, 2\sqrt{3} \pi ,0\right), \nonumber \\ 	\vect{b}_3 &= \left(0,\frac{4 \pi}{\sqrt{3}}, 5 \sqrt{\frac{2}{3}} \pi\right).
\end{align}
For $\h=0$ and isotropic couplings, Weyl points of positive chirality are located at $W_1=5/8 ~\vect{b}_1 \!+\! 3/4 ~ \vect{b}_2\! +\! 3/8 ~\vect{b}_3$ and  $W_2=-1/8 ~\vect{b}_1 \!+\! 1/4 ~ \vect{b}_2\! +\! 1/8 ~\vect{b}_3$
and due to particle-hole symmetry, a pair of negative charge is found at $W_3\!=\!-W_2$ and $W_4\!=\!-W_1$. 

\begin{figure}
	\includegraphics{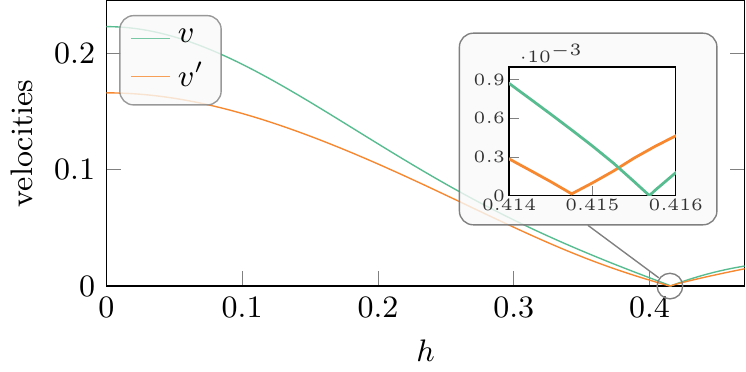}
	\caption{\textbf{Evolution of the velocities \boldmath and \boldmath{$v'$}} describing the linear dispersion around the Weyl point in the directions \textbf{n} and \textbf{n'}.}
	\label{fig:weylvelocities}
\end{figure}

Setting $J_z'$ to zero decouples the lattice to an effectively 2D system with normal vector $\vect{n}' = \left( 1,\sqrt{3}, 0 \right)$. As $|J_z'|$ is decreased, the velocity $v'$ describing the linear dispersion in direction $\vect{n}'$ decreases and reaches zero for $J_z'=0$, giving rise to a gapless line. These observations also hold for $J_z$ and the velocity $v$ in direction $\vect{n} = \left(-\sqrt{6}, \sqrt{2}, -5/2 \right)$. For $J_z\!=\!J_z'\!=\!0$, the two planes of zero energy excitations are given by
$5/\sqrt{2}k_x \!- 5/\sqrt{6}k_y \!- 8/\sqrt{3} k_z \!=\! \pm 10\pi /\sqrt{2}$.
Upon increasing the field, $E_z^{\text{eff}}$ and $E_{z'}^{\text{eff}}$ approach zero resulting in the same development of the velocities, as shown in \figref{fig:weylvelocities}. As mentioned in the main text, the two fields where nodal lines in the directions $\vect{n}$ and $\vect{n'}$ appear range very close together.

The field $h$ not only changes $v$ and $v'$ but also moves the Weyl points.
At $h_{c1}$ they are found to be at $W_1=3/4~ \vect{b}_1 \!+\! 3/4 ~ \vect{b}_2 \!+\! 1/2 ~\vect{b}_3$ and $W_2=1/4~ \vect{b}_1 \!+\! 1/4~ \vect{b}_2$, which is the exact same position found for $J_z',h=0$ and arbitrary $J_z$. Their position at $h_{c2}$ is also found in a system with $J_z,h\!=\!0$ and a very small FM $J_z'$.

\subsection{(10,3)b}
\paragraph{Mean-field parameters}
For FM couplings and $\h=0$, the mean-field parameters are
\begin{align}
A_{13} = A_{24} = 0.526, &&
\bar{A}_{23} = \bar{A}_{14} = 1.
\end{align}
They evolve according to \equref{eq:103bmf}.
\paragraph{Stability of the nodal line}
The Dirac points appearing in the honeycomb lattice and the Fermi line in the lattice (10,3)b both have an even codimension and are thus not a priori stable \cite{horavaStabilityFermiSurfaces2005}. For $h\!=\!0$, they are protected by TRS. However, not all TRS-breaking perturbations gap them out, but only those mixing fermions of the same flavor on the same sublattice \cite{burnell_su2_2011}. A magnetic field in [001] direction can not do so by construction and Dirac points and Fermi line remain stable.
It was shown by \citet{nasu_successive_2018}, that the stability is reflected in the existence of an antiunitary symmetry preserving the Bloch matrix. Using the simple form of the mean-fields, this transformation on the honeycomb lattice has the compact form $U = VK$, where $K$ is complex conjugation and $V= \mathbf{1}_2 \otimes \sigma_x$, with the $2\times2$ identity matrix $\mathbf{1}_2$. Owing to this symmetry, their exists a $\mathbb{Z}_2$ invariant protecting the Dirac points \cite{nasu_successive_2018}.
This argument can be expanded to the lattice (10,3)b by considering only two momentum directions as physical and the third one as a tunable parameter. Choosing the latter such that the plane of physical momenta cuts the nodal line gives an effective 2D Brillouin zone containing two Dirac points. 
The further conclusions are analogous to \refref{nasu_successive_2018}, with the matrix $V$ now given by $ \mathbf{1}_2 \otimes \sigma_x \otimes \sigma_x$.

\subsection{(10,3)a}
\paragraph{Mean-field parameters}
For FM couplings, the mean-field parameters for $\h=0$ are given by
\begin{align}
	A_{23} = A_{14} = 0.513, && 
	\bar{A}_{23} = \bar{A}_{14} = -1.	
\end{align}
The relations between the mean-fields also hold for $h\!>\!0$. The magnetization is equal on all lattice sites.
\paragraph{Gapless modes}
Here, we show that a calculation similar to the one presented in  \secref{sec:103b} can also be done for the lattice (10,3)a.
First, we examine the pure Kitaev model with $J_x\!=\!J_y\!=\!-1$.
Gapless modes appear for
\begin{equation}
	\cos(k_z) = 1 + \frac{J_z^4}{2} + J_z^2 \left( \cos(k_y)-\cos(k_x) \right).
	\label{eq:103a-gapless1}
\end{equation}
For isotropic couplings and finite fields, defining $\xi = -E_z^\text{eff} / E_z^0$, as discussed in the main text,  
the gapless condition becomes
\begin{equation}
\cos(k_z) = 1 + \frac{\xi^4}{2}  + \xi^2 \left( \cos(k_y) - \cos(k_x) \right).
\end{equation}
Again, we find that $\xi$ is zero at $h_c$ and tuning $J_z$ or increasing $h$ deforms the Fermi surface in the same way.

\end{document}